\documentclass[showpacs,preprintnumbers,amsmath,amssymb,showkeys]{revtex4}

\usepackage{epsfig}
\usepackage{dcolumn}
\usepackage{bm}
\usepackage [T1]{fontenc}
\usepackage [latin1,utf8]{inputenc}
\usepackage{graphicx}
\usepackage{amsmath}
\usepackage{amsxtra}
\usepackage{color}
\usepackage[spanish,english]{babel}
\usepackage{wrapfig}
\usepackage{amssymb,amsmath}
\usepackage{makeidx}
\usepackage{graphics}
\usepackage{psfrag}
\usepackage{epsfig}
\usepackage{bm}
\usepackage{float}
\usepackage{lineno}
\usepackage{hyperref}
\usepackage{slashed}
\usepackage{rotating}
\usepackage{subfigure}
\usepackage{bm}
\usepackage{float}
\usepackage{multirow}
\usepackage{comment}
\modulolinenumbers[5]
\newcommand{\dbar} {\ensuremath{\, \mathchar'26\mkern-12mu d}}

\begin{document}
	
\title{Thermodynamics of noncommutative quantum Kerr Black Holes}
	
\author{L. F. Escamilla-Herrera}
\altaffiliation[Current address: ]{Instituto de Ciencias Nucleares, Universidad Nacional Aut\'onoma de M\'exico, AP 70543, 04510, M\'exico, DF, M\'exico.}
\email{lenin.escamilla@correo.nucleares.unam.mx} 
\author{E. Mena-Barboza}
\email{emena@cuci.udg.mx}
\affiliation{%
Centro Universitario de la Ci\'enega, Universidad de Guadalajara, Av. Universidad 1115, 47820 Ocotlán Jalisco, M\'exico}%
\author{J.Torres-Arenas}
\altaffiliation[]{}
\email{jtorres@fisica.ugto.mx} 
\altaffiliation[]{}
\affiliation{%
	Divisi\'{o}n de Ciencias e Ingenier\'{\i}as, Campus Le\'{o}n, Universidad de Guanajuato, Loma del Bosque No. 103 Col. Lomas del Campestre, C.P. 37150, Le\'{o}n, Guanajuato, M\'{e}xico}

\date{\today}
	
\begin{abstract}
Thermodynamic formalism for rotating black holes, characterized by noncommutative and quantum corrections, is constructed. From a fundamental thermodynamic relation, equations of state and thermodynamic response functions are explicitly given and the effect of noncommutativity and quantum correction is discussed. It is shown that the well known divergence exhibited in specific heat is not removed by any of these corrections. However, regions of thermodynamic stability are affected by noncommutativity, increasing the available states for which some thermodynamic stability conditions are satisfied. 
\end{abstract}
	
	

\maketitle

\section{Introduction}\label{intro}


It is known that classical thermodynamics formalism can be applied to explore the physical entropy of a black hole, using semiclassical approaches to general relativity. The best known of these approximations was proposed by Bekenstein and Hawking in order to solve the so-called information problem~\cite{Hawking,Hawking2,Bekenstein,Bekenstein1}. They found that the area $A$ of the event horizon of black holes, in an asymptotically flat spacetime, obey a simple relation which is the mathematical analogue of the corresponding entropy for a black hole $S_{BH}$. The results obtained by Bekenstein and Hawking are supported by  Quantum Field Theory in Curved Spacetime formalism. This formalism is not able to give a bound for the precision with which distance measurements are made, such bound presumably must exist, given by the Planck length. One possible way to introduce this bound is by means of noncommutativity of spacetime. In the context of gravity, noncommutativity it is usually introduced using the Seiberg-Witten map, gauging some appropriate group~\cite{Chamseddine}. More recently, a new proposal to introduce an effective noncommutativity is considered, deforming the minisuperspace in cosmological models instead of the spacetime manifold, in such a way that its coordinates do not commute~\cite{Garcia-Compean}. In this work, the latter formalism will be considered in order to take into account noncommutativity for black holes.

Thermodynamics of black holes has a long history, in particular related to the problem of thermodynamic stability. Since the seminal work of Gibbons and Hawking~\cite{Gibbons}, it is known that this problem can be extended to black holes in non asymptotically flat spacetimes. They found that thermodynamic information of de Sitter black holes exhibit important differences with respect to black holes on an asymptotically flat spacetime, which was later corroborated in different works \cite{Davies3,Chamblin}. They found that such black holes emit radiation with a perfect blackbody spectrum. Temperature of these black holes is determined by their surface gravity, which is the same result obtained for the asymptotically flat space case. However, in a de Sitter space also exists a cosmological event horizon, which also emits particles with a temperature proportional to its surface gravity. Therefore, the only way that thermal equilibrium can be achieved is when both surface gravities are equal.

Regarding black holes in an AdS manifold, Hawking and Page~\cite{Hawking-Page} found that thermodynamic stability can be achieved. For this manifold, the gravitational potential produces a confinement for particles with nonzero mass, acting effectively as a cavity of finite volume, where the black hole is contained. Moreover, their heat capacity is positive, which also allows a canonical description of the system.
Another motivation to study thermodynamic stability of black holes is, the known relation with dynamical stability of those systems. For an asymptotically flat spacetime, Schwarzschild black holes are thermodynamically unstable but have dynamical stability~\cite{Price}. On the other hand, for AdS spacetimes, it is known that thermodynamic and dynamical stability are closely related~\cite{Gubser,Wald}.

Since we are considering black holes in asymptotically flat spacetime, it seems legitimate to ask if corrections like noncommutativity or semiclassical ones are capable of modifying the thermodynamics of black holes in order to have thermodynamic stable systems. 

In references \cite{Davies,Martinez,Escamilla1} this is postulated to be the fundamental thermodynamic relation for black holes, which contains all thermodynamic information of the system. Under this assumption, its classical thermodynamic formalism is constructed finding that for black holes, thermodynamic structure of the theory resemble magnetic systems instead of fluids.

As mentioned above, it is well known that for an asymptotically flat spacetime, temperature of black holes is proportional to its surface gravity $\kappa$, as $T=\kappa\hbar/2\pi k_Bc$~\cite{Hawking4}; this semiclassical result was the key that leads to the Bekenstein-Hawking entropy, which is related to the area of its event horizon,

\begin{equation}\label{eq01}
S_{BH}=\frac{c^3A}{4G\hbar}.
\end{equation}
The appropriated metric to describe a rotating black hole is the Kerr one, which can be written as,

\begin{equation}\label{eq02}
ds^2=-\Big(1-\frac{2Mr}{\Sigma}\Big)dt^2-\frac{4Mra\sin^2\theta}{\Sigma}dtd\theta+\frac{\Sigma}{\Delta}dr^2+\Sigma d\theta^2+\frac{A\sin^2\theta}{\Sigma}d\phi^2;
\end{equation}
where,$\Sigma=r^2+a^2\cos^2\theta$, $\Delta=r^2-2Mr+a^2$, $A=(r^2+a^2)^2-a^2\Delta\sin^2\theta$ and $a=J/Mc$. The area of the event horizon of a black hole is given by $A=\int_s\sqrt{\det|g_{\mu\nu}|}ds$. Applying for the elements of the metric tensor given in eq.\eqref{eq02}, the resulting area is:

\begin{equation}\label{eq03}
A=8\pi G^2M^2c^{-4}\Bigg[1+\sqrt{1-\frac{c^2J^2}{G^2M^4}}\Bigg].
\end{equation}
Substituting in eq. \eqref{eq01}, the resulting expression is assumed to be a thermodynamic fundamental relation for Kerr black holes; with $U=Mc^2$, the internal energy of the system and $J$ its angular momentum. This relation can be written as~\cite{Davies},

\begin{equation}\label{eq04}
S_{BH}(U,J)=\frac{2\pi k_B}{\hbar c}\Bigg(\frac{GU^2}{c^{4}}+\sqrt{\frac{G^2U^4}{c^{8}}-c^2J^2}\Bigg);
\end{equation}
where $G$ is the universal gravitational constant, $\hbar$ is the reduced Planck constant, $c$ is the speed of light in vacuum and $k_B$ is the Boltzmann constant. We are interested in thermodynamic implications of quantum correction to Bekenstein--Hawking (BH) entropy $S_{BH}$ that have arisen in recent years, in the search for suitable candidates of quantum gravity, namely, the quest for understanding microscopic states of black holes~\cite{Keeler,Sen}; and the inclusion to black hole entropy of noncommutativity. This is given considering that coordinates of minisuperspace are noncommutative~\cite{Obregon}. 
Different corrections to Bekenstein-Hawking entropy have emerged from a variety of approaches in recent years, logarithmic ones are a popular choice among those; arising from quantum corrections to the string theory partition function~\cite{Sen2}. They are related to the low energy or infrared properties of gravity, and are also independent of high energy or ultraviolet properties of the theory~\cite{Keeler,Sen2,Banerjee,Carlip}.
In this work, the selected expression for quantum and noncommutative corrected entropy to work with, which will be denoted as $S^\star$, is obtained according to the ideas presented in \cite{Obregon}. Starting from a diffeomorphism between the Kantowski--Sachs cosmological model which describes a homogeneous but anisotropic universe~\cite{Kantowski}, and the Schwarzschild solution, whose line element inside the event horizon $r<2M$ is given by:

\begin{equation}\label{eq05}
ds^2=-\Bigg(\frac{2M}{t}-1\Bigg)^{-1}dt^2+\Bigg(\frac{2M}{t}-1\Bigg)dr^2+t^2(d\theta^2+\sin^2\theta d\phi^2);
\end{equation}
where the temporal $t$ and the spatial $r$ coordinates swap their role, producing a change on the causal structure of spacetime, i.e., the transformation $t\leftrightarrow r$ is performed and considering the Misner parametrization of the Kantowski--Sachs metric it follows,

\begin{equation}\label{eq06}
ds^2=-N^2dt^2+e^{(2\sqrt{3}\gamma)}dr^2+e^{(-2\sqrt{3}\gamma)}e^{(-2\sqrt{3}\lambda)}(d\theta^2+\sin^2\theta d\phi^2). 
\end{equation} 
In this parametrization $\lambda$ and $\gamma$ play the role of the \textit{cartesian coordinates} in the Kantowski-Sachs minisuperspace.
Comparing eqs. \eqref{eq05} and \eqref{eq06} it is straightforward to note the correspondence between components of the metric tensor, in order to identify the functions $N$, $\gamma$ and $\lambda$. The following step is to obtain the Wheeler DeWitt (WDW) equation for Kantowski--Sachs metric given in \eqref{eq06}, whose parametrization is related with the Schwarzschild solution given in eq. \eqref{eq05}, finding the corresponding Hamiltonian of the system $H$ through the Arnowitt--Deser--Misner (ADM) formalism, to introduce it into the quantum wave equation $H\Psi=0$, where $\Psi(\gamma,\lambda)$ is the wave function. This process leads to the WDW equation whose solution  can be found by separation of variables. 

We are interested in the solution that can be obtained when the symplectic structure of minisuperspace is modified, making the coordinates $\lambda$ and $\gamma$ obey the commutation relation $[\lambda,\gamma]=i\theta$, where $\theta$ is the noncommutative parameter; this relation strongly resembles noncommutative quantum mechanics.
The aforementioned deformation can be introduced in terms of a Moyal product, modifying the original phase space, similarly to noncommutative quantum mechanics~\cite{Gamboa}: 

\begin{displaymath}
f(\lambda,\gamma)\star g(\lambda,\gamma)=f(\lambda,\gamma)e^{\frac{i\theta}{2}[\overleftarrow{\partial}_\lambda\overrightarrow{\partial}_\gamma-\overleftarrow{\partial}_\gamma\overrightarrow{\partial}_\lambda]}g(\lambda,\gamma).
\end{displaymath}
These modifications allow us to redefine the coordinates of minisuperspace in order to obtain a noncommutative version of the WDW equation, 

\begin{equation}\label{eq07}
\Bigg[\frac{\partial^2}{\partial\gamma^2}-\frac{\partial^2}{\partial\lambda^2}+48e^{(-2\sqrt{3}\lambda+\sqrt{3}\theta P_\gamma)}\Bigg]\Psi(\lambda,\gamma)=0;
\end{equation}
where $P_\gamma$ is the momentum on coordinate $\gamma$. The above equation can be solved by separation of variables to obtain the corresponding wave function~\cite{Garcia-Compean}:

\begin{equation}\label{eq08}
\Psi(\lambda,\gamma)=e^{i\sqrt{3}\nu\gamma}K_{i\nu}\Big[4e^{(-\sqrt{3}(\lambda+\sqrt{3}\nu\theta/2)}\Big];
\end{equation}
where $\nu$ is the separation constant and $K_{i\nu}$ are the modified Bessel functions. It can be noticed that in eq. \eqref{eq08} the wave function has the form, $\Psi(\lambda,\gamma)=e^{i\sqrt{3}\nu\gamma}\Phi(\lambda)$; therefore, dependence on the coordinate $\gamma$ is the one of a plane wave. It is worth to mention that this contribution vanishes when thermodynamic observables are calculated.

With the above wave equation for the noncommutative Kantowski--Sachs cosmological model, we are able to derive a modified noncommutative version of the entropy. To that purpose, the Feynman--Hibbs procedure is considered in order to calculate the partition function of the system ~\cite{Feynman}. 
In this approach the separated differential equation for $\lambda$ ,

\begin{equation}\label{eq09}
\Bigg[-\frac{d^2}{d\lambda^2}+48e^{-2\sqrt{3}\lambda+3\nu\theta}\Bigg]\Phi(\lambda)=3\nu^2\Phi(\lambda);
\end{equation}
is considered, and the exponential in the potential term $V(\lambda)=48\exp[-2\sqrt{3}\lambda+3\nu\theta]$ of this equation is expanded up to second order in $\lambda$ and with a change of variables, the resulting differential equation can be compared with the corresponding equation for a quantum harmonic oscillator in one dimension which a non-degenerate quantum system. In the Feynmann-Hibbs procedure the potential is modified by the quantum effects, which in the case of the harmonic oscillator is given by:

\begin{displaymath}
U(x)=V(\bar{x})+\frac{\beta\hbar^2}{24m}V''(\bar{x});
\end{displaymath}
where $\bar{x}$ is the mean value of $x$ and $V''(\bar{x})$ stand for the second derivative of the potential. For the considered changes of variables, the noncommutative quantum corrected potential can be written as,

\begin{equation}\label{eq10}
U(x)=\frac{3}{4\pi}\frac{E_p}{l_p^2}e^{3\nu\theta}\Bigg[x^2+\frac{\beta l_p^2E_p}{12}\Bigg].
\end{equation}
The above potential allow us to calculate the canonical partition function of the system,

\begin{equation}\label{eq11}
Z(\beta)=A\int_{-\infty}^{\infty}e^{-\beta U(x)}dx;
\end{equation}
where $\beta^{-1}$ is proportional to the Bekenstein--Hawking temperature and $A=[2\pi l_p^2E_p\beta]^{-1/2}$ is a constant. Substituting $U(x)$ into \eqref{eq11} and performing the integral over $x$, the partition function is given by:

\begin{equation}\label{eq12}
Z(\beta)=\sqrt{\frac{2\pi}{3}}\frac{e^{3\nu\theta/2}}{E_p\beta}\exp\Bigg[-\frac{\beta^2E_p^2}{16\pi}e^{3\nu\theta}\Bigg];
\end{equation}
through this partition function it is possible calculate any thermodynamic observable, by means of the usual thermodynamic relations for the internal energy and the Legendre transformation for Helmholtz free energy,

\begin{displaymath}
\langle E\rangle=-\frac{\partial}{\partial\beta}\ln Z(\beta); \quad \frac{S}{k_B}=\ln Z(\beta)+\beta\langle E\rangle.
\end{displaymath}
With the equation for internal energy, it is possible to determine the value of $\beta$ as a function of the Hawking temperature $\beta_H=8\pi Mc^2/E_p$, obtaining:

\begin{equation}\label{eq13}
\beta=\beta_He^{-3\nu\theta}\Bigg[1-\frac{1}{\beta_He^{-3\nu\theta}}\frac{1}{Mc^2}\Bigg];
\end{equation}
and with the aid of this relation and the Legendre transformation for Helmholtz free energy along with the partition function $Z(\beta)$, the  entropy for the noncommutative quantum corrected black hole can be found:

\begin{equation}\label{eq14}
S^\star=S_{BH}e^{-3\nu\theta}-\frac{1}{2}k_B\ln\Big[\frac{S_{BH}}{k_B}e^{-3\nu\theta}\Big]+\mathcal{O}(S^{-1}_{BH}e^{-3\nu\theta}).
\end{equation}
The functional form of noncommutative quantum black hole entropy $S^\star$ is basically the same than quantum corrected one, besides the addition of multiplicative factor $e^{-3\nu\theta}$ to Bekenstein--Hawking entropy. For the sake of simplicity, we denote the noncommutative term in this expression as:

\begin{displaymath}
\Gamma=\exp[-3\nu\theta]. 
\end{displaymath}
Through the rest of this paper, natural units: $G=\hbar=k_B=c=1$ will be considered. 

In this work, the previous result found for the Schwarzschild noncommutative black hole will be extended to the rotating black hole case. This extension it is not straightforward, since it requires to obtain an analog expression for the noncommutative entropy of the rotating black hole, through the application of a diffeomorphism between the Kerr metric and some appropriated cosmological model and (for instance) the procedure presented in reference~\cite{Obregon}, to our knowledge, the implementation of this procedure has not been reported. However, would be interesting to have and expression to study the effect of angular momentum on the physical properties of the system.  Therefore, in order to have an \textit{approximated} relation for the extended Kerr black hole entropy, the usual Bekenstein--Hawking entropy for rotating black holes, presented in eq. \eqref{eq04} and the generalized one found in equation \eqref{eq14} will be used to obtain an entropy for rotating black holes including quantum and noncommutative effects. Starting from the fact that eq. \eqref{eq14} is correct, whatever be the expression for the non-approximated entropy for the quantum noncommutative Kerr black hole, it is clear that our proposed entropy will be a good approximation for small values of $J$ when compared to the values of $U^2$. Hence, the corrected entropy that we will analyze is:	

\begin{equation}\label{eq15}
S^\star=2\pi\Gamma\Big(U^2+\sqrt{U^4-J^2}\Big)-\frac{1}{2}\ln\Big[2\pi\Gamma\Big(U^2+\sqrt{U^4-J^2}\Big)\Big].
\end{equation}

In the following, all the thermodynamic expressions with superindex $\star$ will stand up for the noncommutative quantum corrected quantities derived from corresponding $S^\star$ entropy, and quantities without subindexes or superindexes will represent their noncommutative Bekenstein--Hawking counterparts. It is known from observational data that noncommutative parameter in spacetime is small~\cite{Aoki,Joby}; however, for entropy $S^\star$, noncommutativity on the coordinates of minisuperspace is considered instead. It is expected such parameter to be small as well~\cite{Sabido}, nonetheless, actual bounds of $\theta$ are not well known yet.

\begin{figure}[t!]
	\centering
	\subfigure[]{\includegraphics[width=0.45\textwidth]{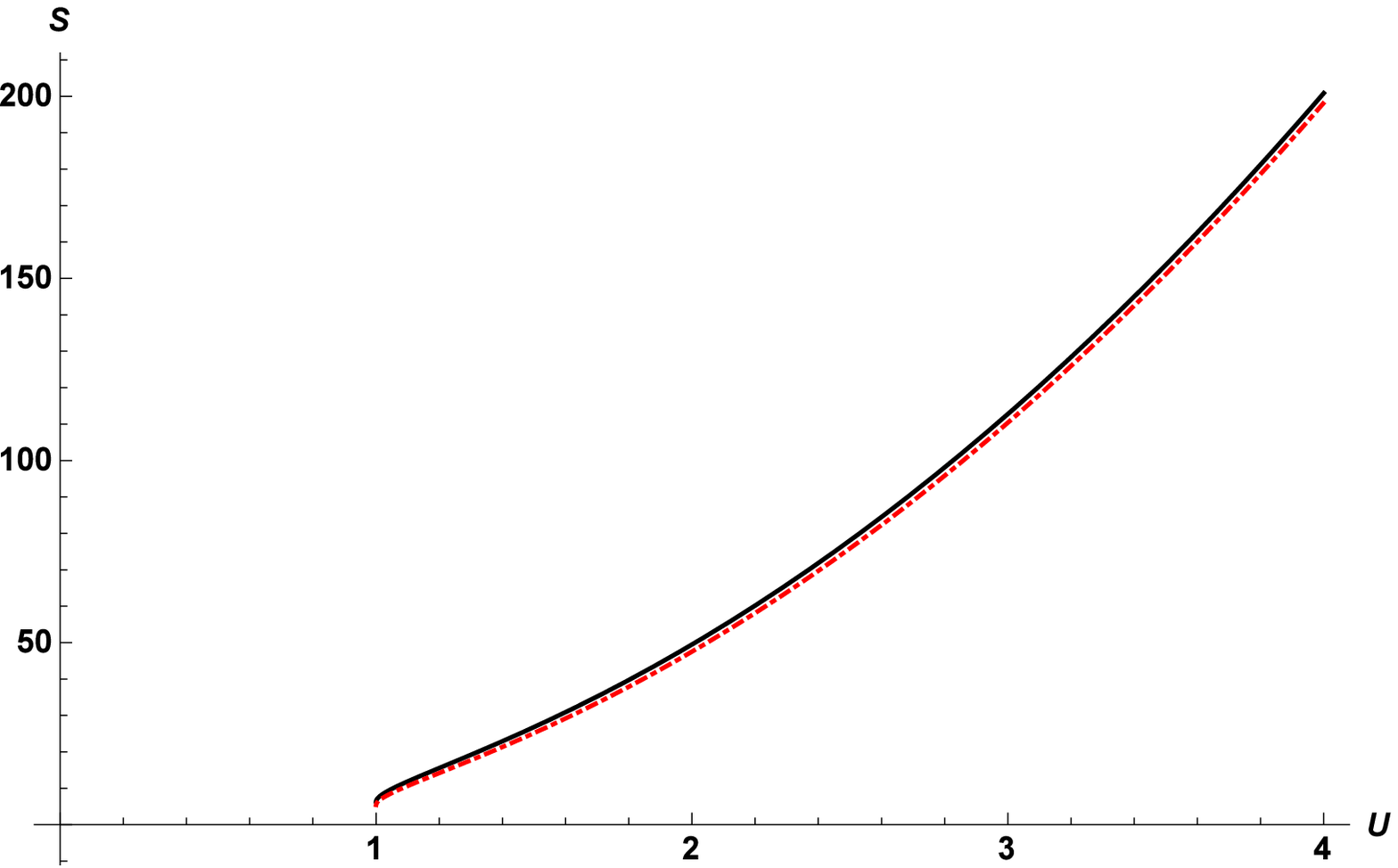}}
	\subfigure[]{\includegraphics[width=0.45\textwidth]{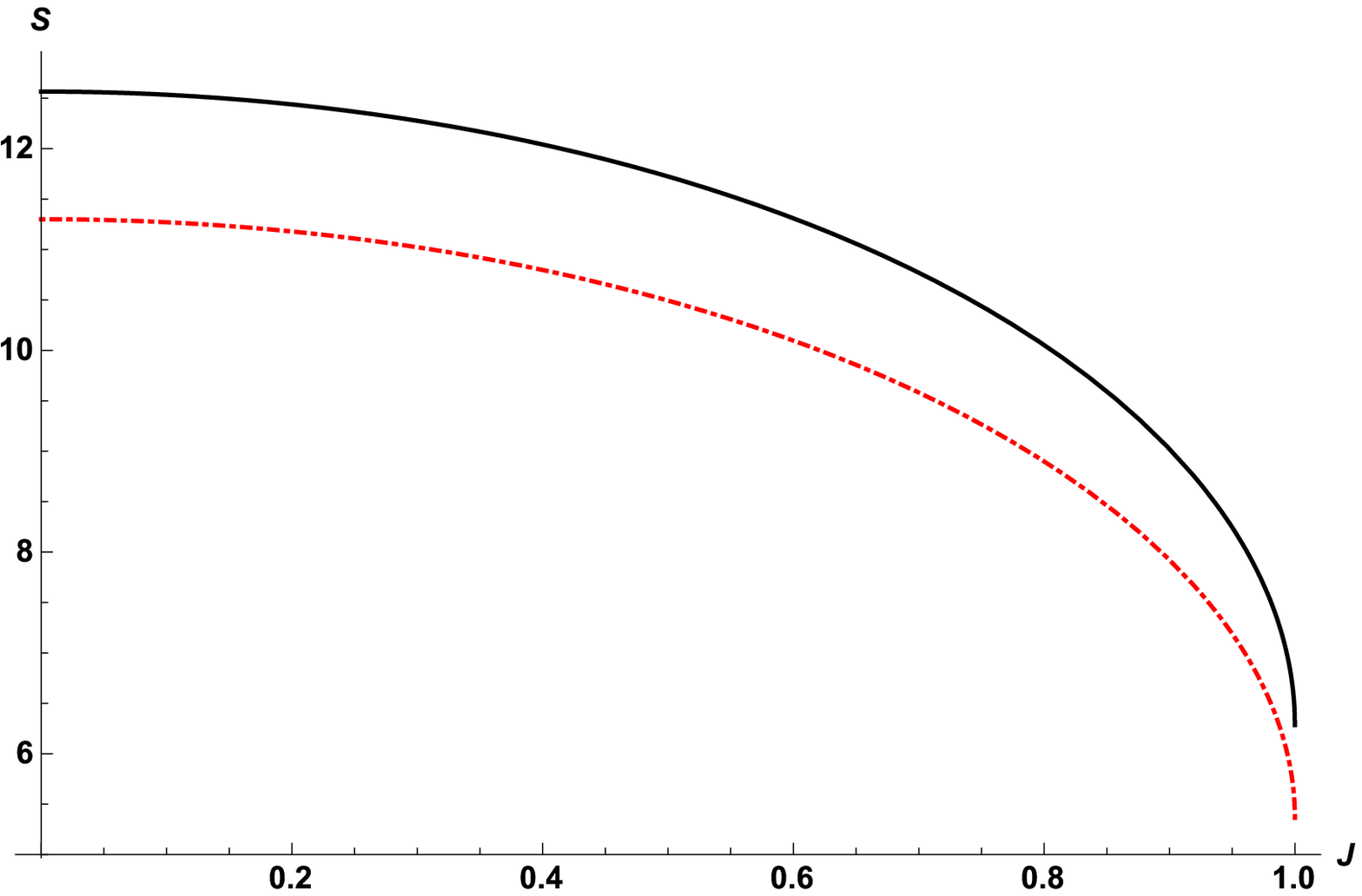}}
	\caption{Comparison between Bekenstein--Hawking (solid line) and quantum corrected (dash-dot line) entropies. (\textbf{a}) Entropy as a function of internal energy for $J=1$, $S=S(U,1)$. (\textbf{b}) Entropy as a function of angular momentum for $U=1$, $S=S(1,J)$. \label{fig01}}
\end{figure}

In this work, the parameter $\Gamma$ will be considered to be bounded in the interval given by $0<\Gamma\le1$.
As mentioned above, eq. \eqref{eq15} will be assumed as a fundamental thermodynamic relation for Kerr black holes when noncommutative and quantum corrections are considered. It is well known from classical thermodynamics that fundamental thermodynamic relations, contain all thermodynamic information of the system under study~\cite{Callen}, as a consequence, modifications on thermodynamic information originated by the introduced corrections to entropy, are carried through all thermodynamic quantities.\\
In Fig. \ref{fig01} curves for Bekenstein--Hawking and quantum corrected entropy are presented considering only commutative relations ($\Gamma=1$). Fig. \ref{fig01}(a) considers plots for $S=S(U)$ and $S^\star=S^\star(U)$. Bekenstein-Hawking entropy is above of quantum correction one, including the region of small energy where entropy is thermodynamically stable~\cite{Escamilla1}. Fig. \ref{fig01}(b) shows the same entropy as functions of angular momentum instead, for fixed values of $U$; it can be noticed that Bekenstein--Hawking entropy is above $S^\star$ in all of considered dominion as well. 
A similar analysis can be performed over noncommutativity, finding that for small values of $\theta$, variations over $S$ and $S^\star$ are negligible.

The goal of this manuscript is to explore how these considerations introduced into Bekenstein--Hawking entropy, change thermodynamic information contained in this new fundamental relation, in particular, thermodynamic stability and the existence of thermodynamic phase transition for these systems.
In the following an outline of this work is presented. Section \ref{sec:1} examines the different thermodynamic equations of state  and their behavior when considering aforementioned modifications to entropy. The same analysis is carried out in section \ref{sec:2}, considering thermodynamic response functions instead. In section \ref{sec:3}, a discussion of thermodynamic stability and phase transitions for Kerr black holes is presented. Some conclusions of this work are given in section \ref{sec:4}.

\section{Equations of state}\label{sec:1}

Fundamental Bekenstein--Hawking thermodynamic relation in entropic representation for Kerr black holes has the form $S_{BH}=S_{BH}(U,J)$. The role of thermodynamic equations of state for Kerr black holes is played by partial derivatives of entropy, $T\equiv(\partial_SU)_J$ and $\Omega\equiv(\partial_JU)_S$ where $\Omega$ is the angular velocity of the black hole and $T$ is its temperature, the following relations in entropic representation are defined:

\begin{equation}\label{eq16}
\frac{1}{T}\equiv\Big(\frac{\partial S_{BH}}{\partial U}\Big)_J; \quad \frac{\Omega}{T}\equiv-\Big(\frac{\partial S_{BH}}{\partial J}\Big)_U.
\end{equation} 
These definitions are also valid for quantum corrected entropy $S^\star$.
Explicit equations of state in entropic representation, $T^\star=T^\star(U,J)$ and $\Omega^\star=\Omega^\star(U,J)$ for noncommutative quantum corrected entropy can be written as,

\begin{subequations} \label{eq17}
	\begin{align}
	\frac{1}{T^{\star}}=\frac{U\Big(4\pi\Gamma\sqrt{U^4-J^2}+4\pi\Gamma U^2-1\Big)}{\sqrt{U^4-J^2}}; \\
	\frac{\Omega^{\star}}{T^{\star}}=\frac{1}{2}\frac{J\Big(4\pi\Gamma\sqrt{U^4-J^2}+4\pi\Gamma U^2-1\Big)}{\sqrt{U^4-J^2}\Big(U^2+\sqrt{U^4-J^2}\Big)}.
	\end{align}
\end{subequations}
In addition, for nonconmutative Bekenstein--Hawking entropy the corresponding equations of state are given by,

\begin{subequations} \label{eq18}
	\begin{align}
	\frac{1}{T}=\frac{4\pi\Gamma U\Big(U^2+\sqrt{U^4-J^2}\Big)}{\sqrt{U^4-J^2}},\\
	\frac{\Omega}{T}=\frac{2\pi\Gamma J}{\sqrt{U^4-J^2}}.
	\end{align}
\end{subequations}

\begin{figure}[t!]
	\centering
	\subfigure[]{\includegraphics[width=0.45\textwidth]{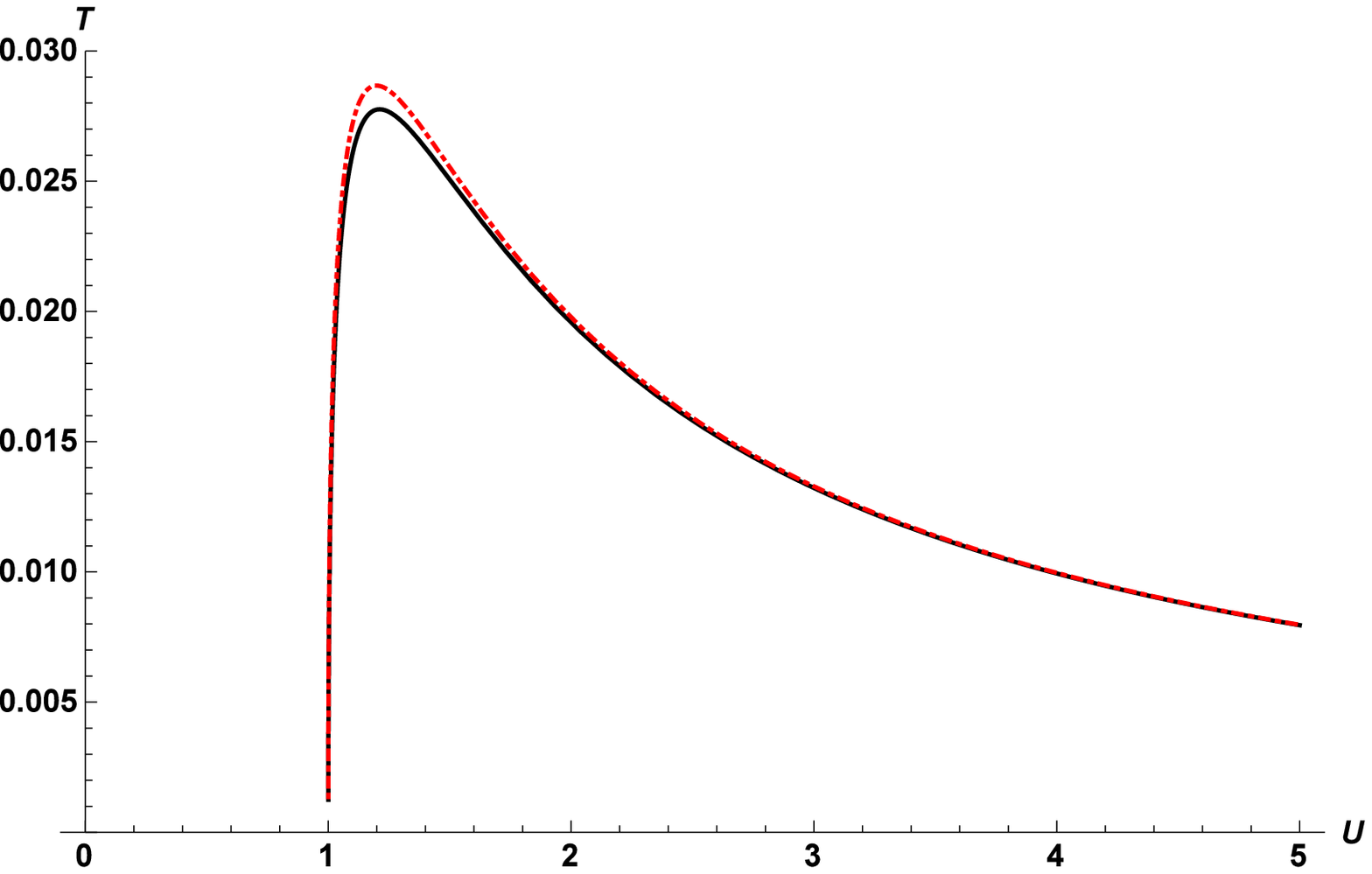}}
	\subfigure[]{\includegraphics[width=0.45\textwidth]{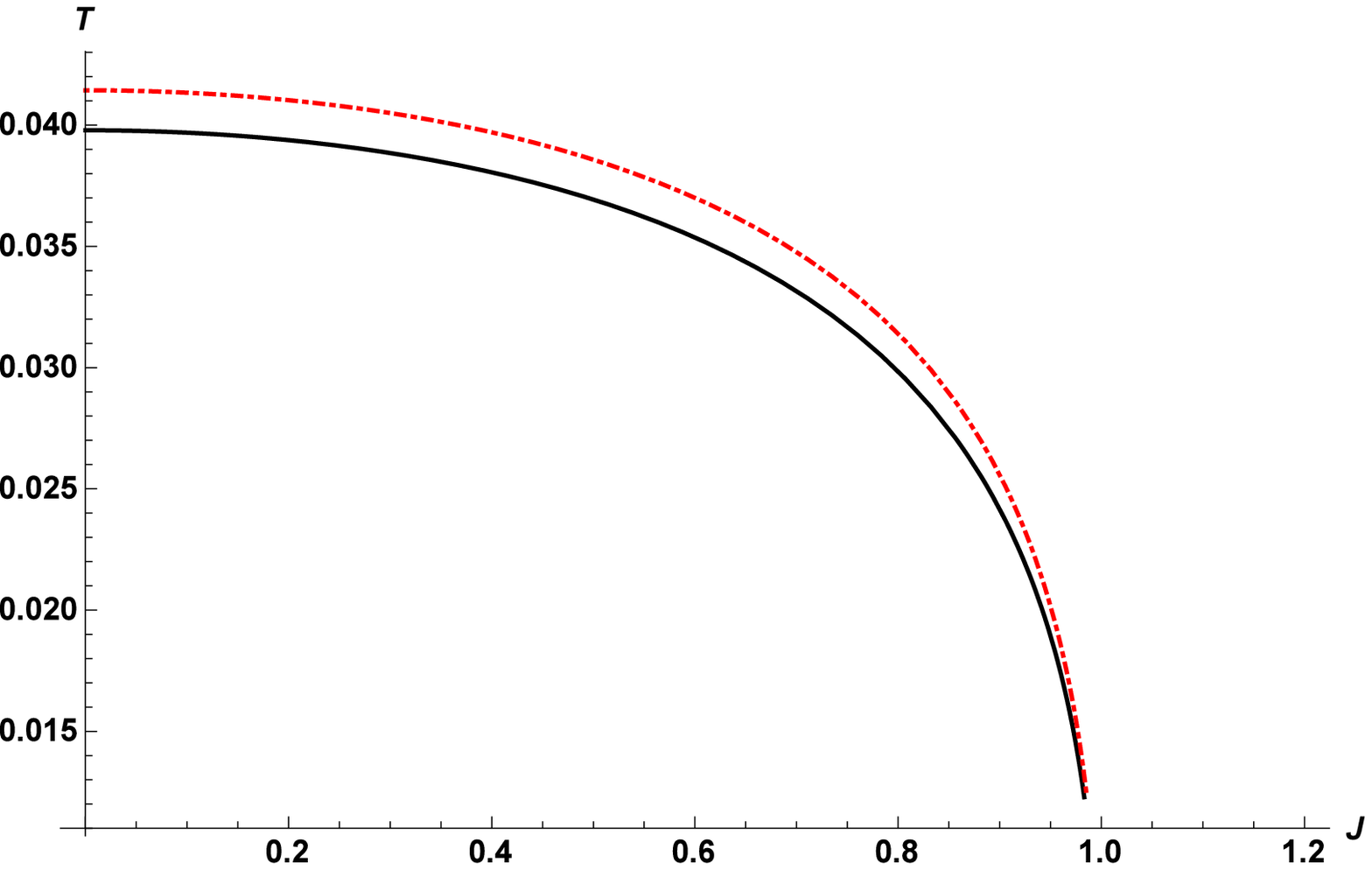}}
	\caption{Plots of Bekenstein--Hawking and quantum corrected temperatures for $\Gamma=1$. (\textbf{a}) $T(U,1)$ (solid) vs $T^\star(U,1)$ (dash-dot) as a function of internal energy considering $J=1$. (\textbf{b}) The same plots of temperature for variations in angular momentum at $U=1$. \label{fig02}}
\end{figure}

The overall effect of noncommutativity over $T$ and $T^\star$ was analyzed, considering different values of $\Gamma$, including the commutative case ($\Gamma=1$). A noticeable effect of this parameter over these curves can be found; nonetheless, it does not change functional behavior either of $T$ or $T^\star$. In order to illustrate how the introduced quantum correction affect thermodynamic properties of black holes when compared with Bekenstein--Hawking ones, a graphical comparison between $T$ and $T^\star$ is performed in Fig. \ref{fig02} for $\Gamma=1$. As expected, Bekenstein-Hawking curves are very similar to those obtained through corrected entropy. Nevertheless, for temperature it is possible to remark that $T^\star(U,J)$ is slightly higher than $T(U,J)$, which is the opposite result that the one obtained for entropy, indicating that variations of entropy for a given change in its internal energy are greater for quantum corrected entropy when compared to Bekenstein--Hawking one.

\begin{figure}[t!]
	\centering
	\subfigure[]{\includegraphics[width=0.45\textwidth]{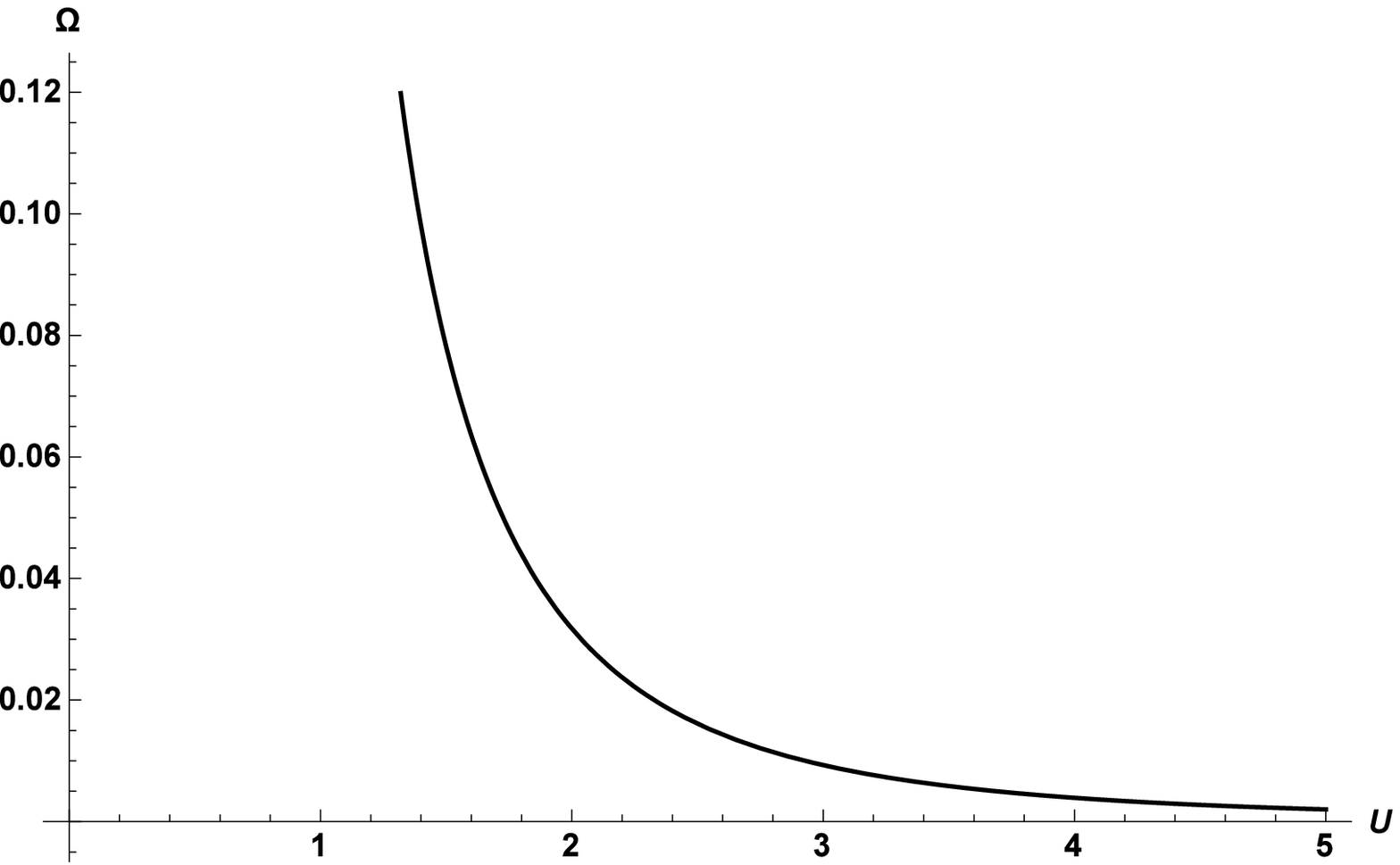}}
	\subfigure[]{\includegraphics[width=0.45\textwidth]{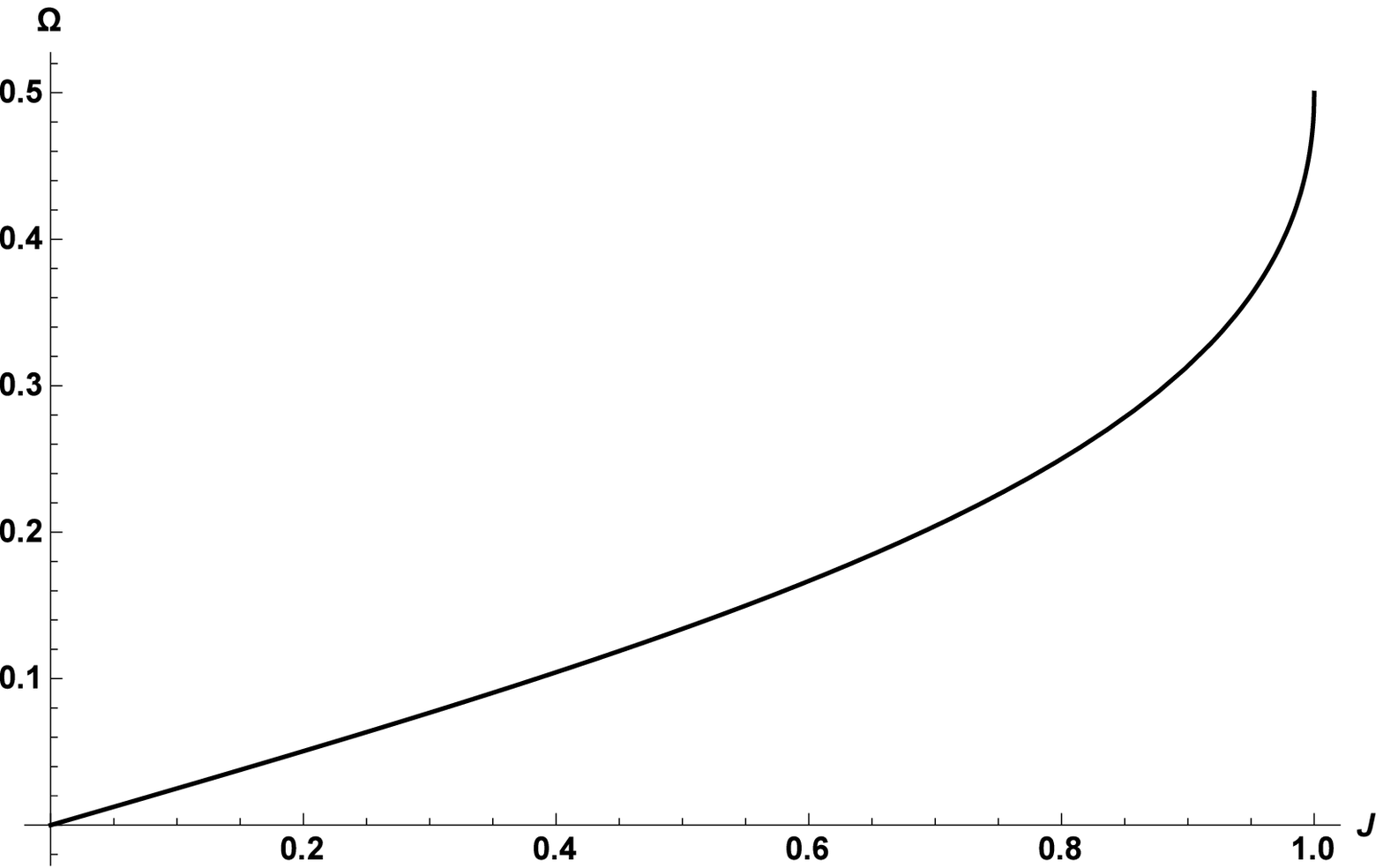}}
	\caption{Plots of angular velocity for both Bekenstein--Hawking and quantum corrected entropies. (\textbf{a}) $\Omega$ as function of internal energy for $J=1$. (\textbf{b}) Angular velocity as a function of angular momentum considering $U=10$. \label{fig03}}
\end{figure}

It was mentioned above that when considering values in the vicinity of $\Gamma=1$, temperature is minimally affected by noncommutativity. Smaller values of $\Gamma$ were also tested, as a consequence of this consideration, maximum values capable to reach by $T$ and $T^*$ are no\-ti\-cea\-ble increased. It must be remarked that changing this parameter does not alter the shape of the curves.\\
Regarding the angular velocity, it is an interesting result to remark that this property is independent of both quantum and nonconmutative corrections to entropy, namely,

\begin{equation}\label{eq19}
\Omega=\Omega^\star=\frac{J}{2U\Big(U^2+\sqrt{U^4-J^2}\Big)}.
\end{equation} 
In Fig. \ref{fig03} angular velocity in entropic representation is presented. Fig. \ref{fig03}(a) shows $\Omega$ as a function of energy for $J=1$; in this case $\Omega$ increases until it reaches a maximum value from which it becomes complex, and it is determined by square root that appears in the denominator of eq. \eqref{eq19}.

\section{Results and Discussion}

\subsection{Response functions}\label{sec:2}

\begin{figure}[t!]
	\centering
	\subfigure[]{\includegraphics[width=0.45\textwidth]{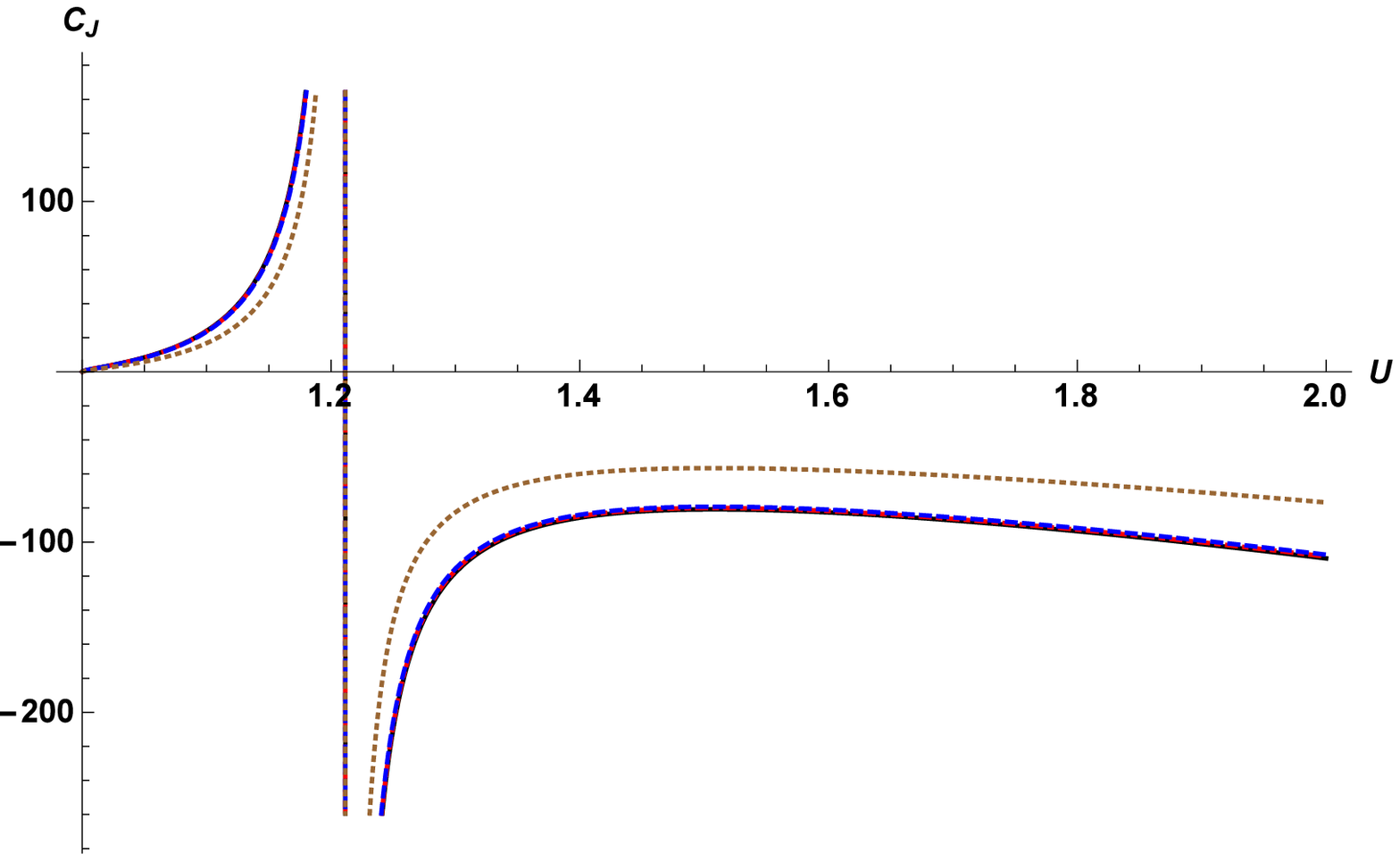}}
	\subfigure[]{\includegraphics[width=0.45\textwidth]{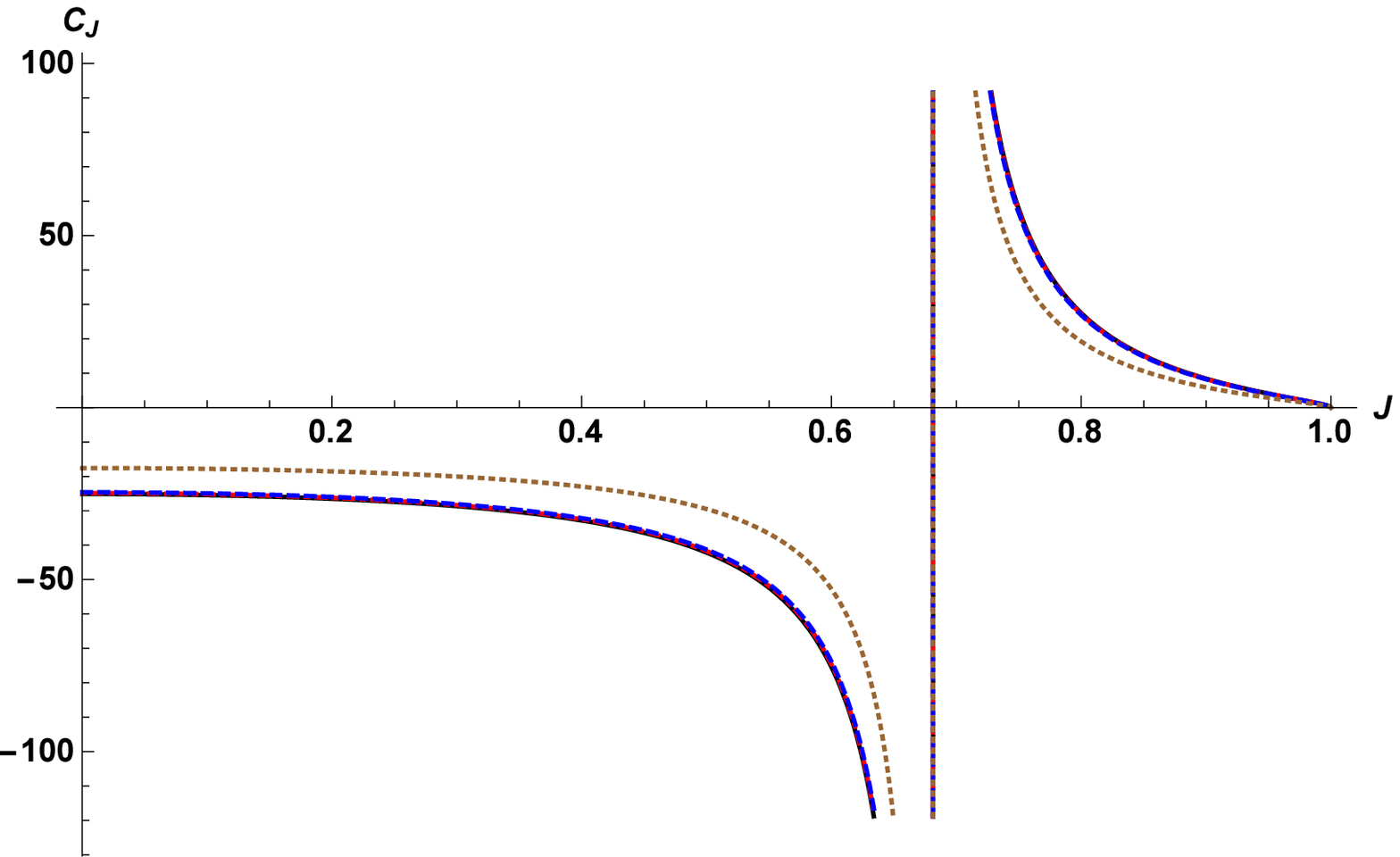}}
	\caption{Specific heat capacity at constant $J$ for Kerr black hole considering different values of $\Gamma$ exhibiting a discontinuity, the following values of noncommutativity parameter are considered: $\Gamma=1$ (solid), $\Gamma=0.99$ (dashed-dot), $\Gamma=0.98$ (dashed) and $\Gamma=0.7$ (dotted). (a) $C_J$ is presented as a function of energy at $J=1$. (b) Plots of specific heat as a function of angular momentum for $U=1$. \label{fig04}}
\end{figure}

Response functions contain valuable information about the thermodynamic behavior of systems; therefore, this topic must be addressed in order to study the changes if any, introduced to black hole thermodynamic properties by noncommutativity and quantum correction to its entropy. 
It is possible to define thermodynamic response functions for a Kerr black hole considering Bekenstein--Hawking entropy, following the structure exhibited by magnetic systems in such a way that resulting $TdS$ equations~\cite{Stanley} are completely analogous to their magnetic counterparts~\cite{Escamilla1}. Subsequently and following this resemble with magnetic systems, thermodynamic response functions are defined in this work without any weight factor except for heat capacities, defined with such factor given by inverse of temperature.
Even with the above considerations, thermodynamic response functions for black holes can also be constructed following the structure commonly associated with fluids, as made in Refs.~\cite{Davies,Bagher}.\\
The first response to be analyzed is the heat capacity at constant angular momentum, defined as:

\begin{equation}\label{eq20}
C_J\equiv\Big(\frac{\dbar Q}{dT}\Big)_J=T\Big(\frac{\partial S}{\partial T}\Big)_J=\Big(\frac{\partial U}{\partial T}\Big)_J;
\end{equation}
in entropic representation $T=T(U,J)$, which makes convenient to write,

\begin{equation}\label{eq21}
C_J=\Big(\frac{\partial T}{\partial U}\Big)^{-1}_J.
\end{equation}
Corresponding heat capacity at constant angular momentum for noncommutative Bekenstein--Hawking and quantum corrected entropies are respectively given by:

\begin{subequations} \label{eq22}
	\begin{align}
	C_J= \frac{4\pi\Gamma U^2\sqrt{U^4-J^2}\Big(U^2+\sqrt{U^4-J^2}\Big)}{\Big(U^4+J^2-2U^2\sqrt{U^4-J^2}\Big)};\\
	C^\star_J=\frac{U^2\sqrt{U^4-J^2}\Big[4\pi\Gamma\sqrt{U^4-J^2}+4\pi\Gamma U^2-1\Big]^2}{\Big[4\pi\Gamma U^6+U^4-12\pi\Gamma U^2J^2+4\pi\Gamma\sqrt{U^4-J^2}\Big(U^4-J^2\Big)+J^2\Big]}.
	\end{align}
\end{subequations}
In Fig. \ref{fig04}, $C_J$ is plotted as a function of internal energy for a given angular momentum in Fig. \ref{fig04}(a), and as a function of $J$ for $U=10$ in Fig. \ref{fig04}(b). The most relevant feature exhibited in those plots is the divergence that appears in both curves, which divides heat capacity in two regions, one  where $C_J$ is positive and another where specific heat becomes negative. It is well known for some time, that black holes display divergences in response functions, specifically in heat capacity~\cite{Davies}. This feature also has been found to appear in high dimensional black hole models~\cite{Bagher,Banerjee2,Mansoori,Lala}, and often has been related with phase transitions in black holes, this topic will be discussed and presented later. 

For response functions, lower values of $\Gamma$ were also tested. Specifically in Fig. \ref{fig04}, it can be observed that varying this parameter do not change the appearance of discontinuity of $C_J$. Although this divergence is not removed by noncommutativity,  negative part of specific heat is reduced as $\Gamma$ is reduced.
Divergence in $C_J$ can be traced to the roots of denominator in eq. (\ref{eq22}.a):

\begin{displaymath}
U^4+J^2-2U^2\sqrt{U^4-J^2}=0,
\end{displaymath}
this function has one real positive root for $J$,

\begin{equation}\label{eq23}
J_{\text{sing}}=\sqrt{-3+2\sqrt{3}}\cdot U^2\approx0.68U^2.
\end{equation} 
Which is the same value found in Ref.~\cite{Davies}, for a Kerr black hole with Bekenstein--Hawking entropy. It must be remarked that the above root it is not affected by noncommutativity corrections.

To locate this divergence across different equations of state in entropic representation, it is necessary to substitute the above value $J_{\text{sing}}$ in each equation of state. For example, for the angular velocity equation of state given in eq. (\ref{eq18}b), if eq. \eqref{eq23} is substituted into this expression, it leads to:

\begin{equation}\label{eq24}
\frac{\Omega}{T}\approx5.83\Gamma,
\end{equation}
or $\Omega\approx5.8271\Gamma T$. Therefore, in the plane $\Omega$--$T$ there is a straight line which divides this plane in two regions where $C_J>0$ before the divergence, and the second one where $C_J<0$. Analogously, for Bekenstein--Hawking entropy and temperature, the corresponding function which divides the $S$--$U$ plane is given by the parabola determined by:

\begin{equation}\label{eq25}
S\approx10.88\Gamma U^2;
\end{equation}
additionally, plane $T$--$U$ is divided by the following straight line,

\begin{equation}\label{eq26}
T\approx29.73\Gamma U.
\end{equation} 

\begin{figure}[t!]
	\centering
	\subfigure[]{\includegraphics[width=0.45\textwidth]{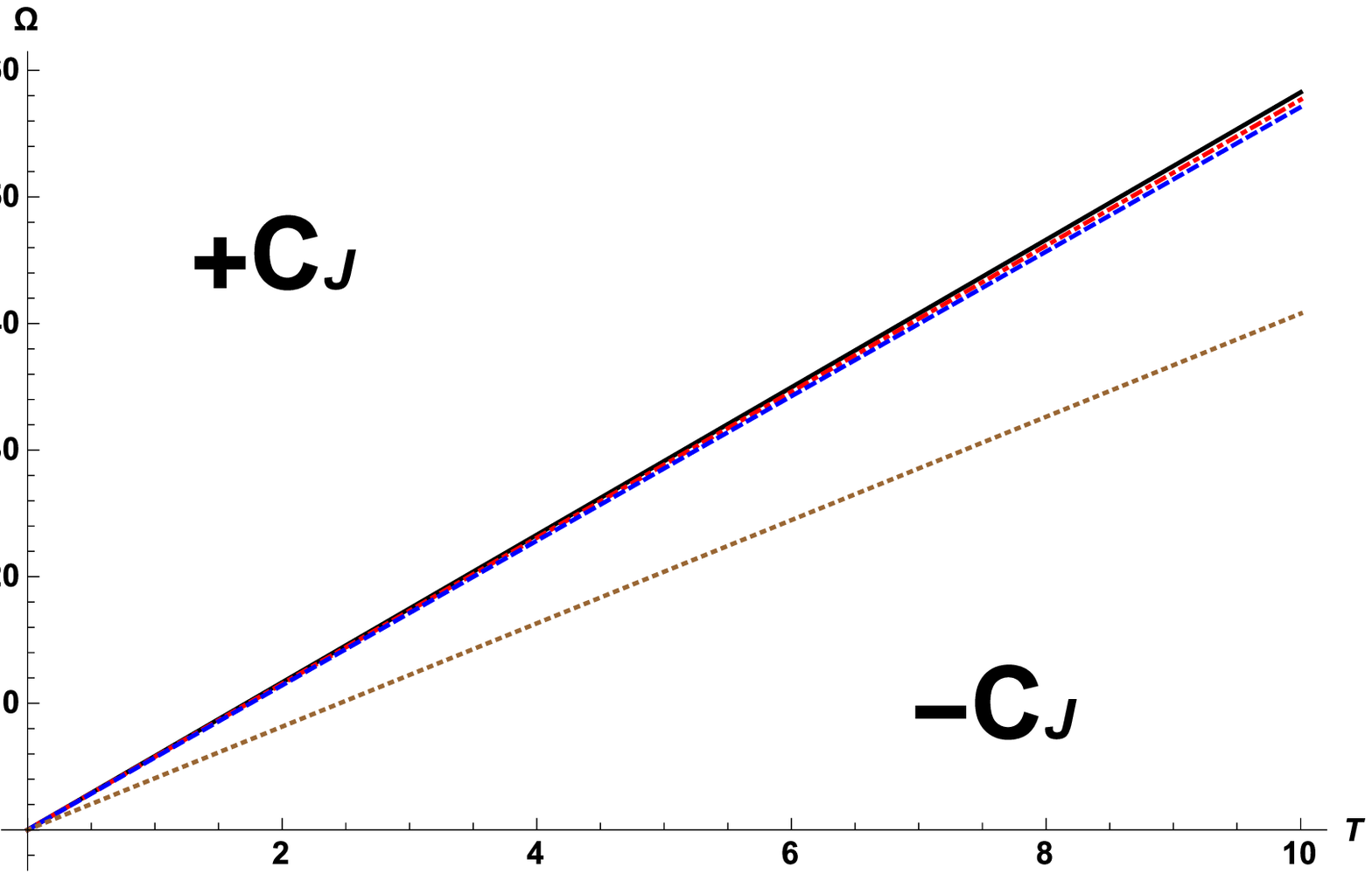}}
	\subfigure[]{\includegraphics[width=0.45\textwidth]{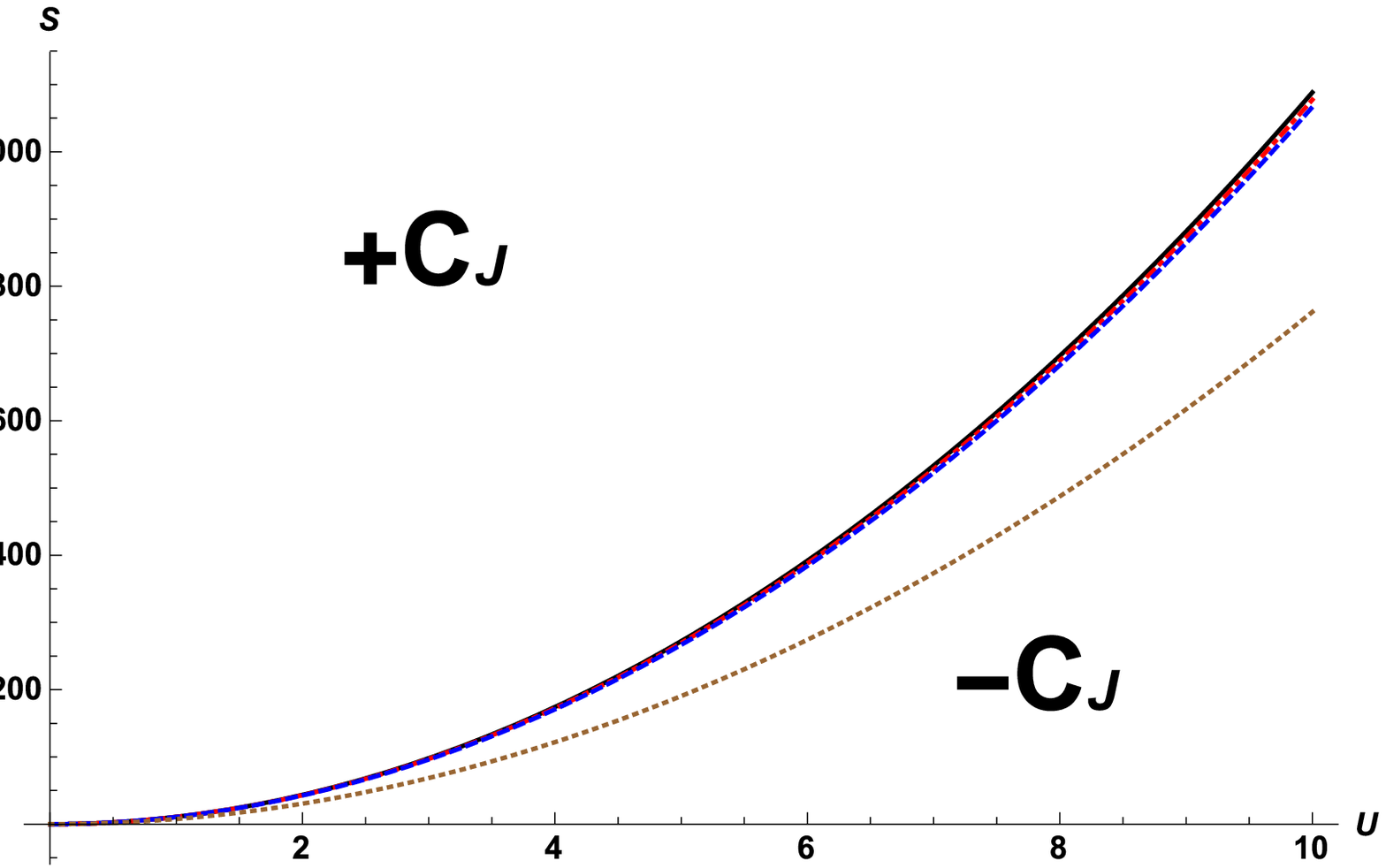}}
	\subfigure[]{\includegraphics[width=0.45\textwidth]{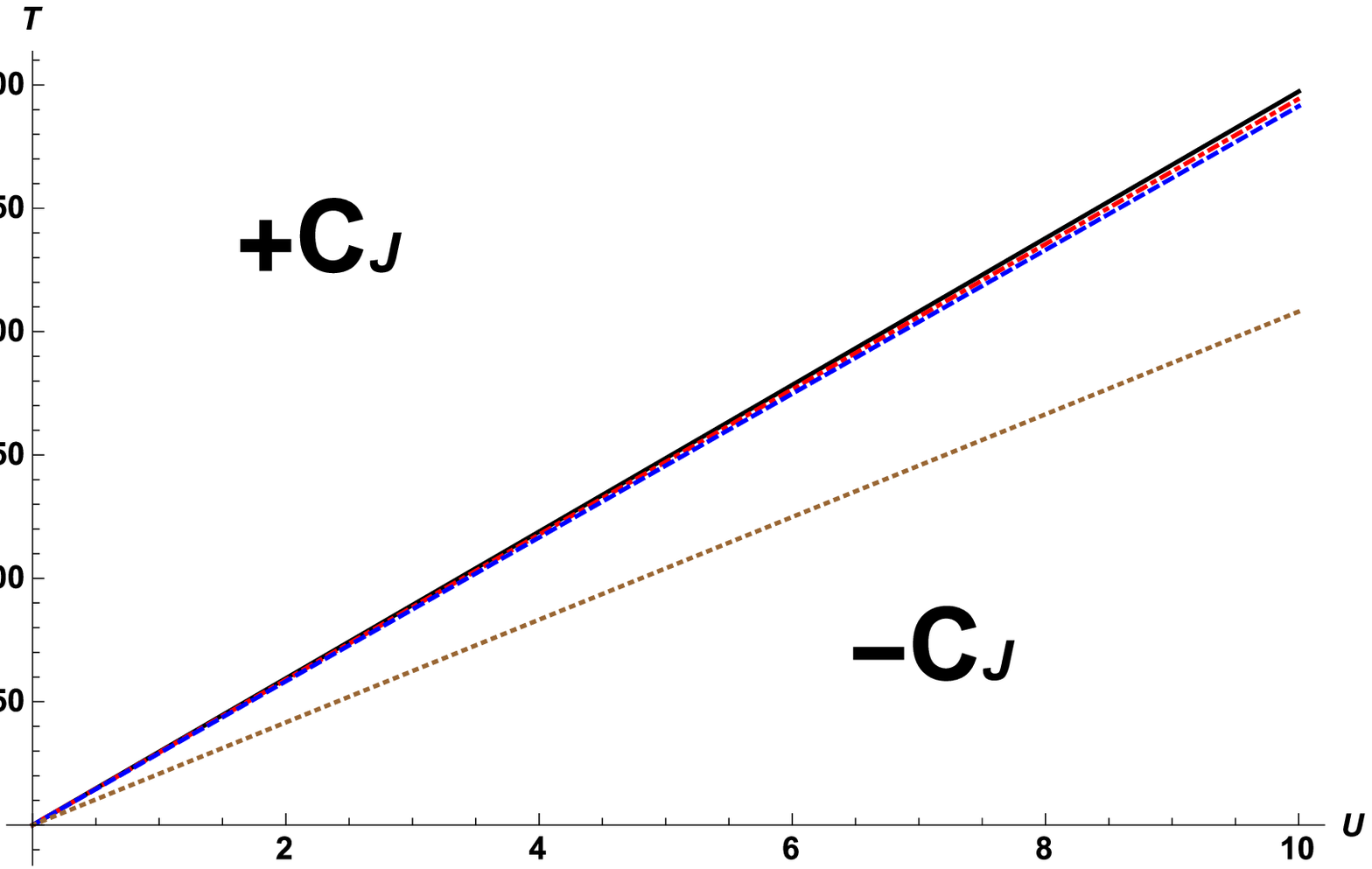}}
	\caption{Thermodynamic planes considering different values of $\Gamma$ showing location of divergence found in specific heat, $\pm C_J$ indicates the sign of this response function in each region; in this thermodynamic planes, the following values of $\Gamma$ are presented: $\Gamma=1$ (solid), $\Gamma=0.99$ (dashed-dot), $\Gamma=0.98$ (dashed) and $\Gamma=0.7$ (dotted). (\textbf{a}) Plane $\Omega$--$T$ divided by a line representing discontinuity given in eq. \eqref{eq23}. (\textbf{b}) Plane $S$--$U$ shows a parabola dividing regions where $C_J$ is positive or negative. (\textbf{c}) Plane $T$--$U$ depicts another straight line separating both regions. \label{fig05}}
\end{figure}

One highlight from eqs. \eqref{eq24}--\eqref{eq26} is that all of them are linear functions of noncommutativity parameter $\Gamma$ (and exponentially on $\theta$). In Fig. \ref{fig05} all the above corresponding thermodynamic planes are plotted, changes introduced by noncommutativity in each of those planes increase the area of the region where specific heat $C_J$ is positive, therefore reducing the possible values for which this response function can become negative. 
These changes near $\Gamma=1$ are subtle, but if $\Gamma$ is taken out this neighborhood, the area where $C_J<0$ becomes considerably smaller. 
This is an important result that can be related to thermodynamic stability, and it will be discussed in the next section. It is enough to indicate that noncommutativity modify the region of available thermodynamically stable states for the system.

Regarding the noncommutative quantum corrected specific heat capacity at constant angular momentum $C^\star_J$, results are fairly similar to its Bekenstein--Hawking counterpart, quantum correction do not removes the discontinuity in this response function. Similarly as $C_J$, it can be located by finding the roots in denominator of eq. (\ref{eq22}b),

\begin{equation}\label{eq27}
4\pi\Gamma U^6+U^4-12\pi\Gamma U^2J^2+4\pi\Gamma U^4\sqrt{U^4-J^2}-4\pi\Gamma J^2\sqrt{U^4-J^2}+J^2=0;
\end{equation}
to simplify the above expression a set of manageable functions of $J$ are obtained, by substituting a sequence of values for $U$ into \eqref{eq27}. It is possible to solve each of these function assuming $J>0$, constructing a set of coordinate pairs $(U,J)$. This process is repeated several times to obtain a relevant sample in order to accurately represent the root of eq. \eqref{eq27}.
Straightforward, a plot can be constructed, using the set $(U,J)$ to obtain an expression for the root by least--squares method. When plotted, points clearly exhibit a quadratic behavior, and a fitting process lead us to:
\begin{equation}\label{eq28}
J_{\text{sing}}\approx0.02176+0.68126U^2.
\end{equation}
When compared to  eq. \eqref{eq23}, this result show that discontinuity in $C^\star_J$ is almost the same than the one found for Bekenstein--Hawking specific heat. This result is another indication, as shown above for first--order derivatives, that thermodynamic properties obtained from noncommutative quantum corrected entropy are very close with their noncommutative Bekenstein-Hawking counterparts.

Unlike results presented in eq. \eqref{eq23}, there is not a simple function that can be used to describe  behavior of neither equation of state through its corresponding phase plane for eq. \eqref{eq28}.
The only plane that can be plotted directly is $S^\star$--$U$ one, although its functional behavior is not simple. Correspondingly to $C_J$, exhibited behavior in this plane for quantum corrected expression is very similar to the one obtained for noncommutative Bekenstein-Hawking one, in Fig. \ref{fig05}(b); including the role played parameter $\Gamma$. If $C_J$ and $C^\star_J$ are compared in the commutative case, it is found that  $C_J(U,J)>C^\star_J(U,J)$  by a slight margin in all their dominion. 
Another response function that can be defined for Kerr black holes is the isothermic rotational susceptibility~\cite{Escamilla1},

\begin{equation}\label{eq29}
\chi_T\equiv\Big(\frac{\partial J}{\partial\Omega}\Big)_T.
\end{equation}
Alternative functional forms for this response function can be obtained in entropic representation. With some algebraic manipulation it is possible to write isothermic rotational susceptibility as:

\begin{equation}\label{eq30}
\chi_T=\Bigg[\Big(\frac{\partial\Omega}{\partial J}\Big)_U-\frac{(\partial\Omega/\partial U)_J}{(\partial T/\partial U)_J}\Big(\frac{\partial T}{\partial J}\Big)_U\Bigg]^{-1};
\end{equation} 
it is possible to work in entropic representation with the above result since both equations of state $T=T(U,J)$ and $\Omega=\Omega(U,J)$ are available.

\begin{figure}[t!]
	\centering
	\subfigure[]{\includegraphics[width=0.45\textwidth]{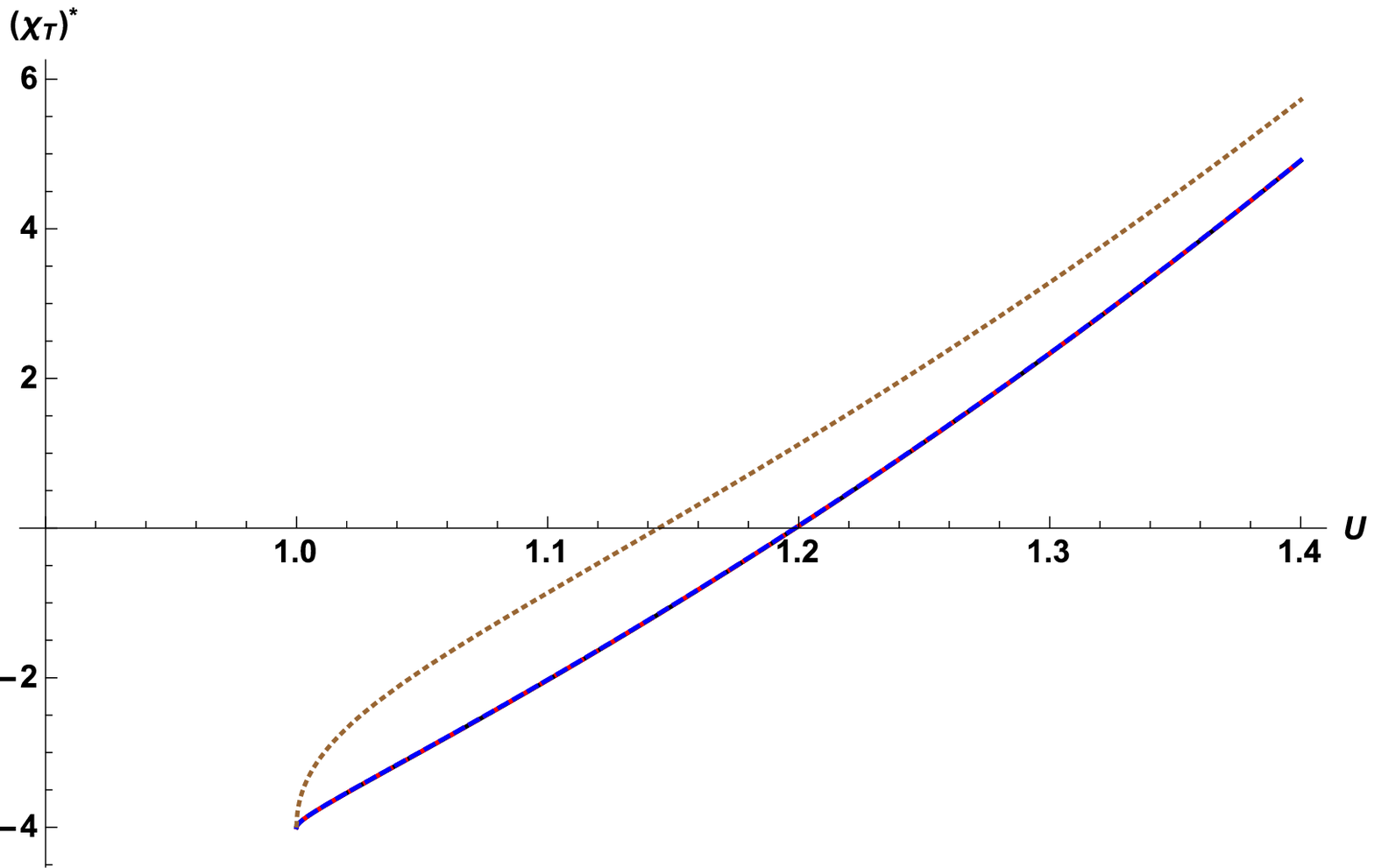}}
	\subfigure[]{\includegraphics[width=0.45\textwidth]{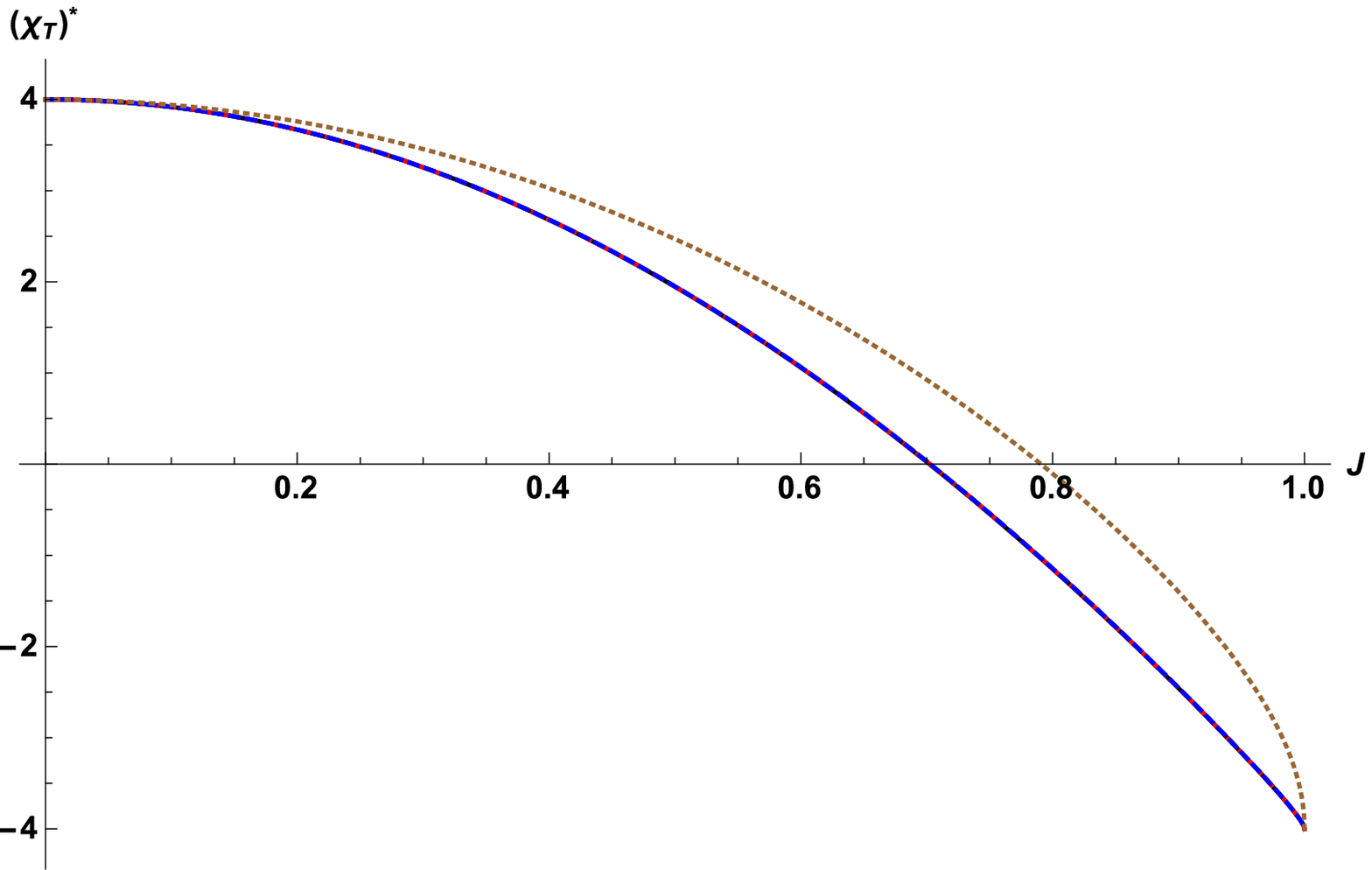}}
	\caption{Noncommutative quantum corrected isothermal rotational susceptibility. (\textbf{a}) $\chi^\star_T$ as function of energy for $J=1$, exhibiting a monotonically growing function, given that noncommutativity does not greatly affect $\chi^\star$, the following values for this parameter are chosen: $\Gamma=1$ (solid), $\Gamma=0.99$ (dashed-dot), $\Gamma=0.98$ (dashed) and $\Gamma=0.2$ (dotted).  (\textbf{b}) Curves for isothermal susceptibility as a function of $J$ considering $U=1$. \label{fig06}}
\end{figure}

For noncommutative Bekenstein--Hawking entropy, isothermic rotational susceptibility can be written as

\begin{equation}\label{eq31}
\chi_T=-\frac{2}{U^3}\Bigg[\Big(U^2+\sqrt{U^4-J^2}\Big)\Big(U^4+J^2-2U^2\sqrt{U^4-J^2}\Big)\Bigg];
\end{equation}
a remarkable feature of this material property is that it is independent of the noncommutative parameter $\Gamma$, in analogy of angular velocity presented in eq.~\eqref{eq19}. 
Furthermore, $\chi_T$ is well defined in all its domain, on the opposite to $C_J$. Plots for $\chi_T$ are presented in Fig. \ref{fig06}, from these curves it can be noted that $\chi_T\to0$ when $J\approx0.68U^2$ or equivalently, $U\approx1.21\sqrt{J}$. $\chi_T(U)$ have a region of negative values, which is also related to thermodynamic stability.

With respect to noncommutative quantum corrected entropy, its corresponding isothermal rotational susceptibility is obtained by application of eq. \eqref{eq30},

\begin{displaymath}
\chi^\star_T=\frac{-4\pi\Gamma U^6-U^4+12\pi\Gamma U^2J^2-J^2-4\pi\Gamma(U^4-J^2)^{3/2}}{-8\pi\Gamma U^8-U^6+4\pi\Gamma U^4J^2+2U^2J^2+\sqrt{U^4-J^2}(-8\pi\Gamma U^6-U^4+J^2)}
\end{displaymath}
\begin{equation}\label{eq32}
\times 2U\Big(U^2+\sqrt{U^4-J^2}\Big)^2;
\end{equation}
this relation is indeed a function of $\Gamma$, and it is presented in Fig. \ref{fig06}.  Effect of noncommutativity in $\chi^\star_T$ is almost nonexistent, and if different values of this parameter near $\Gamma=1$ are plotted together for $\chi^\star_T$ all resulting curves overlap. Only when the vicinity near $\Gamma\to0$ is considered, noncommutativity effect on $\chi^\star_T$ is perceivable. Changes produced by $\Gamma$ in numerator of eq. \eqref{eq32} are countered by its role in denominator of the same expression.\\
$\chi^\star_T$ goes to zero around $J\approx0.7$ for $U=1$, which corresponds to the location of divergence for $C^\star_J$. Negative region still appears in $\chi^\star_T$ as observed in Bekenstein-Hawking isothermal rotational susceptibility. Additionally, if both $\chi_T$ and $\chi^\star_T$ are compared, the latter is always slightly above to the former, namely, $\chi_T(U,J)<\chi^\star_T(U,J)$.
Another thermodynamic response function to be analyzed is the isentropic rotational, defined as~\cite{Escamilla1}:

\begin{equation}\label{eq33}
\chi_S\equiv\Big(\frac{\partial J}{\partial\Omega}\Big)_S;
\end{equation}
similarly to $\chi_T$, it is straightforward to obtain by rewriting in terms of $U$ and $J$,

\begin{equation}\label{eq34}
\chi_S=\Bigg[\Big(\frac{\partial \Omega}{\partial J}\Big)_U+\Omega\Big(\frac{\partial\Omega}{\partial U}\Big)_J\Bigg]^{-1}.
\end{equation}
This relation leads to the same result for both noncommutative Bekenstein--Hawking and quantum corrected entropies since $\Omega=\Omega^\star$, as showed in eq. \eqref{eq19}. Therefore,

\begin{equation}\label{eq35}
\chi_S=\chi^\star_S=4U^3;
\end{equation}
which is well defined in all its dominion. Independence of $\Gamma$ in the above result is a consequence that for both $S$ and $S^\star$, angular velocity is also independent of this parameter.  

\begin{figure}[t!]
	\centering
	\subfigure[]{\includegraphics[width=0.45\textwidth]{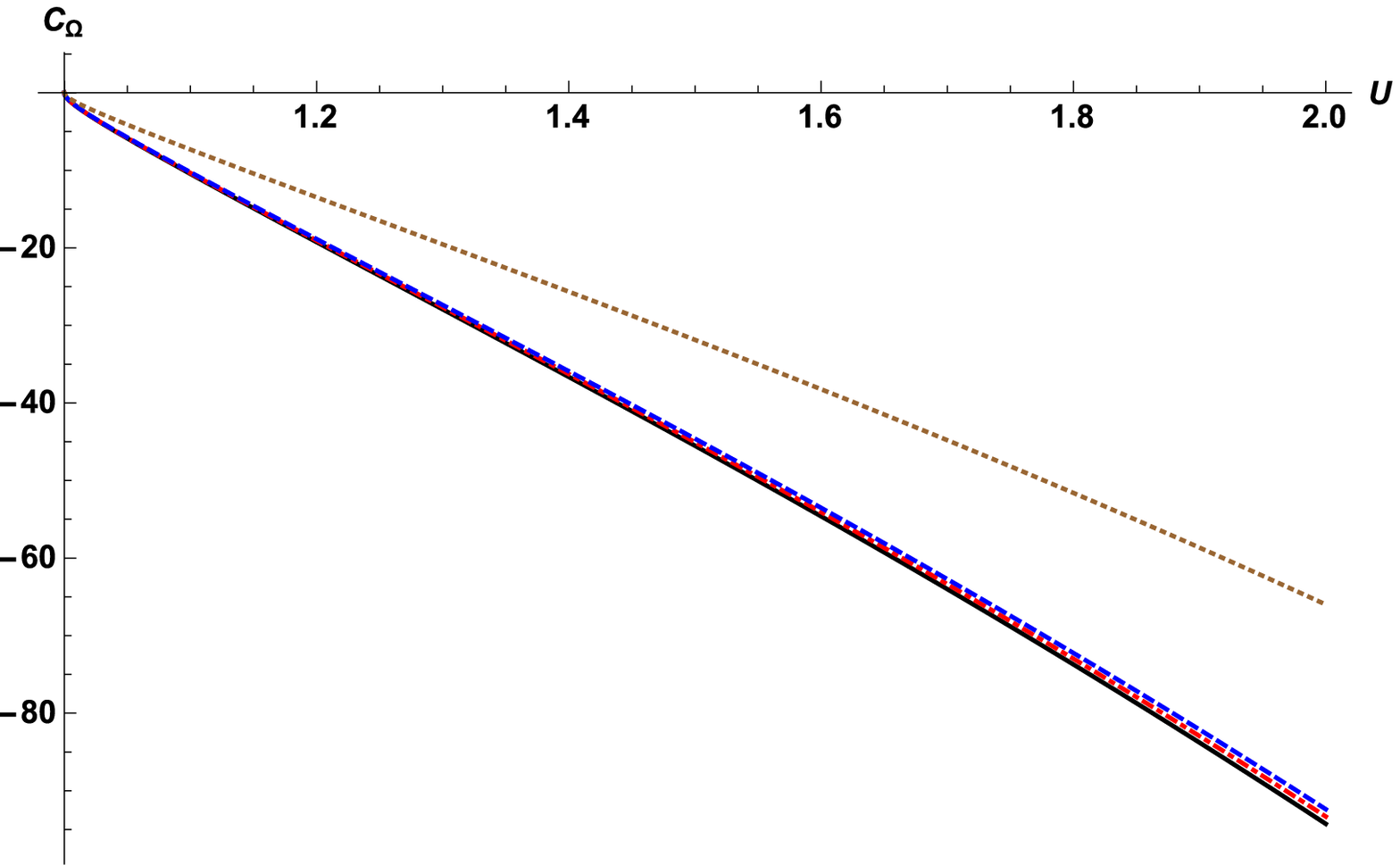}}
	\subfigure[]{\includegraphics[width=0.45\textwidth]{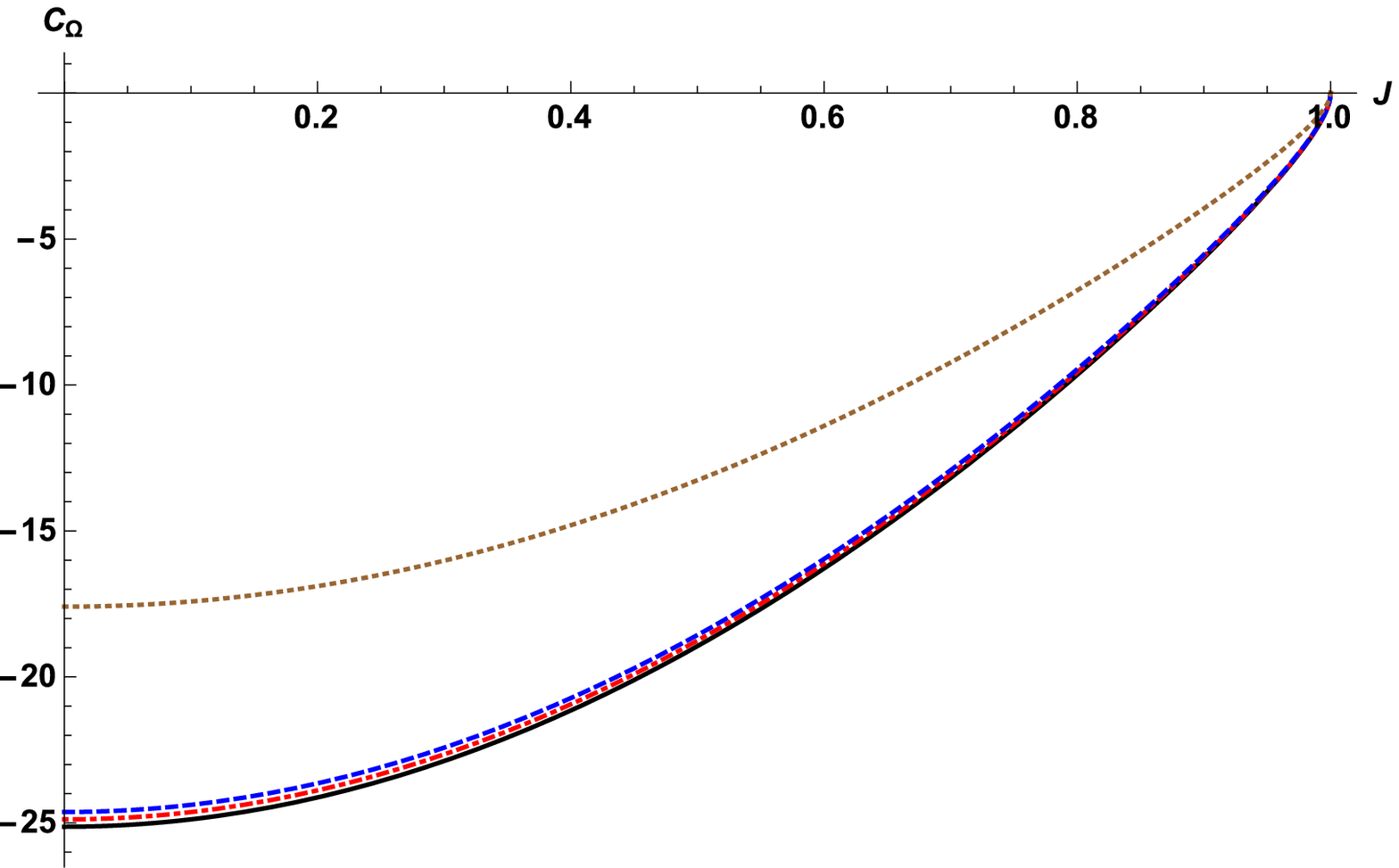}}
	\caption{Specific heat capacity at constant angular velocity for different values of $\Gamma$. (\textbf{a}) $C_\Omega$ is plotted as a function of internal energy for $J=1$; for this response fuctions the following values of $\Gamma$ where considered: $\Gamma=1$ (solid), $\Gamma=0.99$ (dashed-dot), $\Gamma=0.98$ (dashed) and $\Gamma=0.7$ (dotted). (\textbf{b}) Curves of $C_\Omega$ for $U=1$ as a function of angular momentum. \label{fig07}}
\end{figure}

For Kerr black holes, specific heat can also be defined maintaining constant angular velocity,

\begin{equation}\label{eq36}
C_\Omega\equiv\Big(\frac{\dbar Q}{dT}\Big)_\Omega=T\Big(\frac{\partial S}{\partial T}\Big)_\Omega;
\end{equation}
$TdS$ equations provide a set of algebraic relations between response functions that can be applied in order to find analytical expressions for $C_\Omega$ and $C^\star_\Omega$. The following relations between material properties arise~\cite{Escamilla1}: 

\begin{subequations} \label{eq37}
	\begin{align}
	\chi_T(C_\Omega-C_J)=T\alpha^2_\Omega;\\
	C_\Omega(\chi_T-\chi_S)=T\alpha^2_\Omega;\\
	\chi_SC_\Omega=\chi_TC_J.
	\end{align}
\end{subequations}
Where $\alpha_\Omega$ is the coefficient of thermal induced rotation. Heat capacity at constant angular velocity can be obtained directly from eq. (\ref{eq37}c).\\
Therefore, specific heat for noncommutative Bekenstein--Hawking entropy is given by:

\begin{equation}\label{eq38}
C_\Omega=\frac{2\pi\Gamma\sqrt{U^4-J^2}}{U^4}\Big(-2U^2\sqrt{U^4-J^2}-2U^4+J^2\Big),
\end{equation}
this expression is well defined in all the domain of its variables, and has one discontinuity in the trivial case where $U=0$ (or $M=0$). In Fig. \ref{fig07} this response function is presented for different values of $\Gamma$, it can be noticed that $C_\Omega$ is negative in all its domain.
Noncommutativity reduces the negativity of this heat capacity, finding $C_\Omega\to0$ as $\Gamma\to0$. Nevertheless, this response function remains always negative.

Considering noncommutative quantum corrected entropy, specific heat capacity at constant angular velocity can be expressed as:
\begin{equation}\label{eq39}
C^\star_\Omega=-\frac{1}{2}\frac{\sqrt{U^4-J^2}(4\pi\Gamma U^2+4\pi\Gamma\sqrt{U^4-J^2}-1)^2(-2U^4-2U^2\sqrt{U^4-J^2}+J^2)}{-8\pi\Gamma U^8-U^6+4\pi\Gamma U^4J^2+2U^2J^2-(8\pi\Gamma U^6+U^4-J^2)\sqrt{U^4-J^2}};
\end{equation}
its graphical representation  is very similar to the one exhibited by $C_\Omega$. 
Although parameter $\Gamma$ plays a more complicated role in eq. \eqref{eq37}, the overall effect of noncommutativity in $C^\star_\Omega$ leads to a very close behavior to the one observed in $C_\Omega$. A direct comparison between both response functions shows that $C^\star_\Omega(U,J)>C_\Omega(U,J)$ in all their domain.

\begin{figure}[t!]
	\centering
	\subfigure[]{\includegraphics[width=0.45\textwidth]{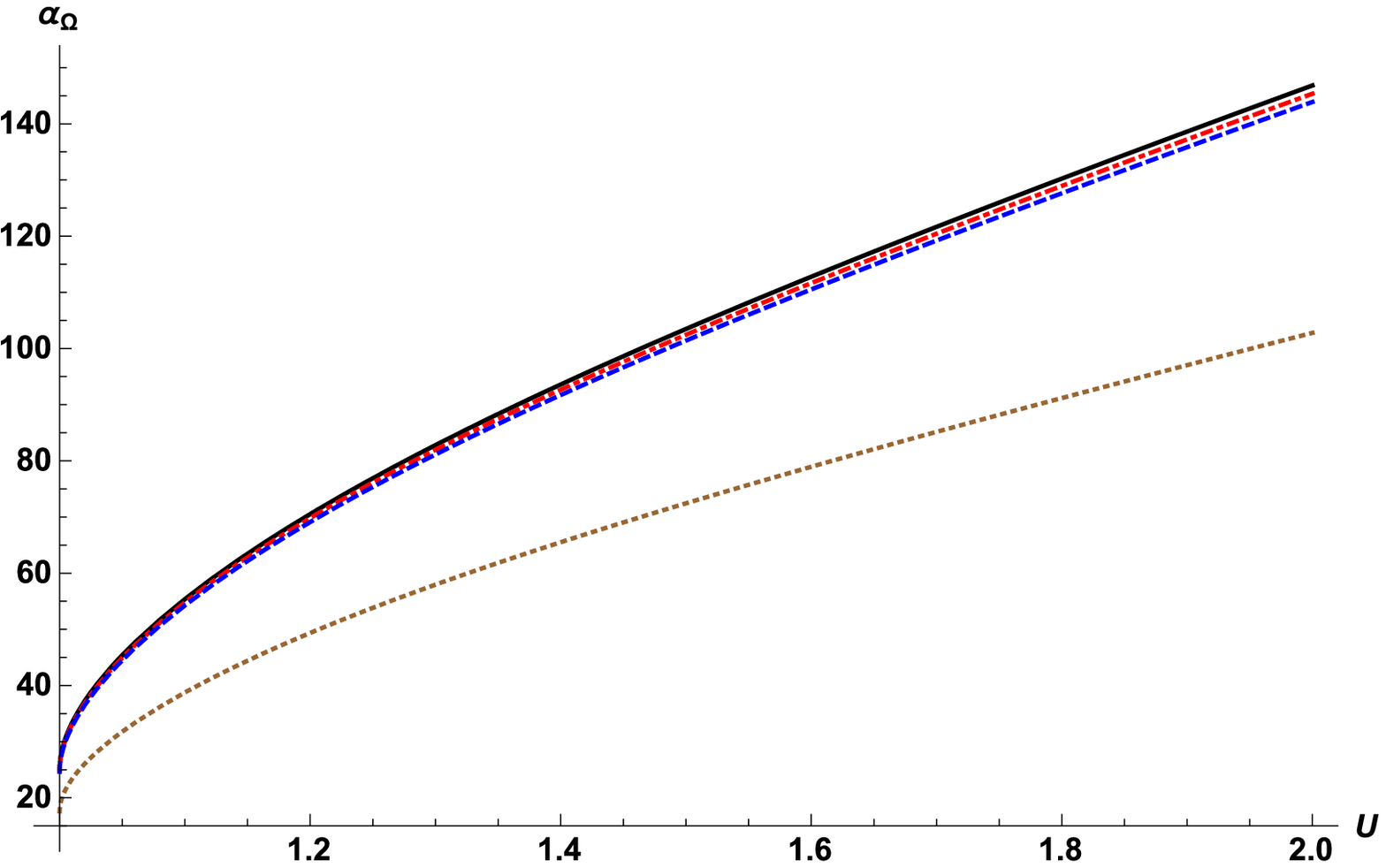}}
	\subfigure[]{\includegraphics[width=0.45\textwidth]{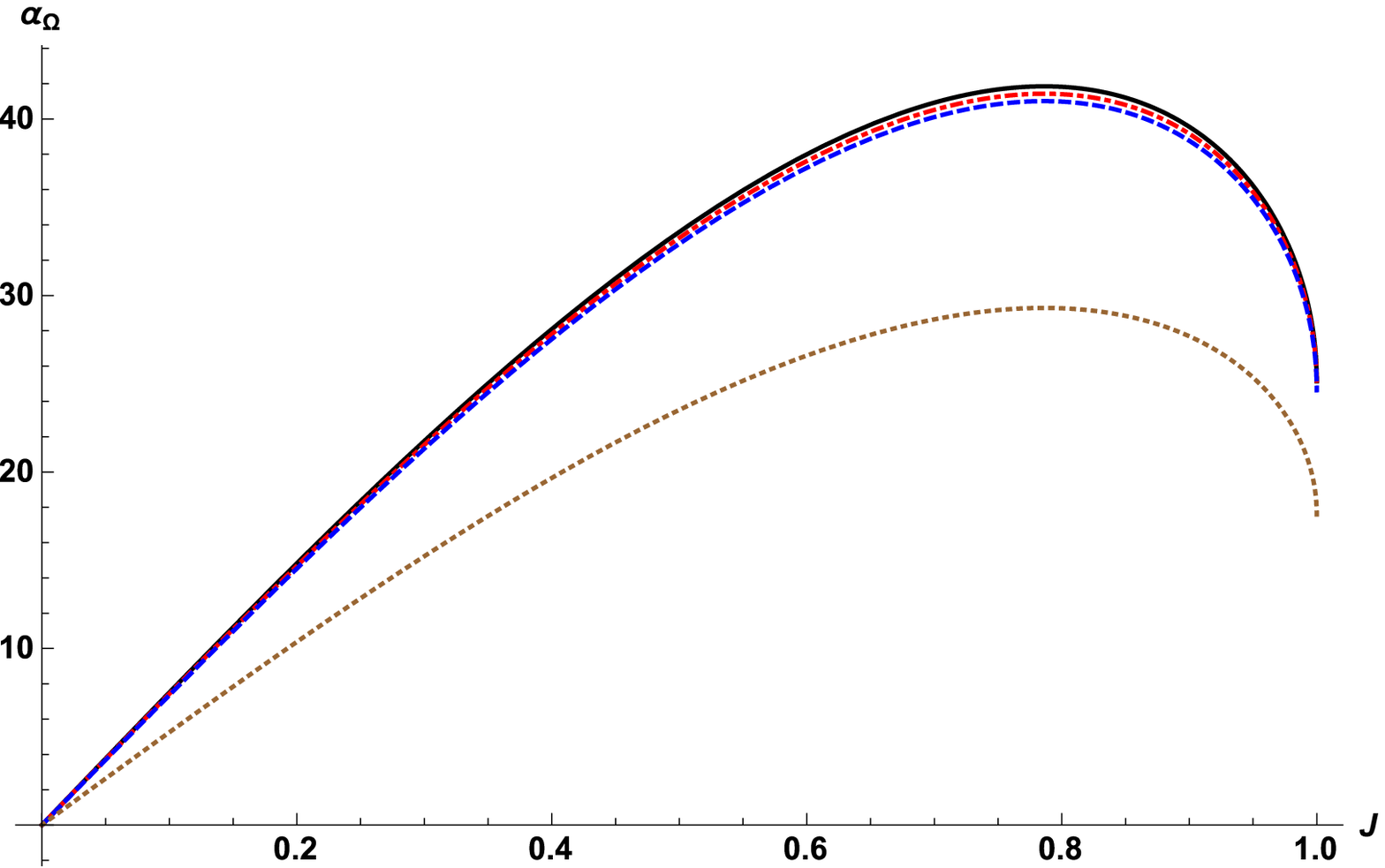}}
	\caption{Coefficient of thermal induced rotation for noncommutative Bekenstein--Hawking entropy. (\textbf{a}) $\alpha_\Omega$ as a function of internal energy, for $J=1$; where the following values of the noncommutativiy parameter are plotted: $\Gamma=1$ (solid), $\Gamma=0.99$ (dashed-dot), $\Gamma=0.98$ (dashed) and $\Gamma=0.7$ (dotted). (\textbf{b}) Plots of the same coefficient varying angular momentum, for internal energy at $U=1$. \label{fig08}}
\end{figure}

The last response function of Kerr black holes studied in this work is the coefficient of thermal induced rotation $\alpha_\Omega$~\cite{Escamilla1},
\begin{equation}\label{eq40}
\alpha_\Omega\equiv\Big(\frac{\partial J}{\partial T}\Big)_\Omega.
\end{equation}
This material property is also calculated indirectly via relations between response functions, either eq. (\ref{eq37}a) or eq. (\ref{eq37}b). For noncommutative Bekenstein--Hawking entropy,
\begin{equation}\label{eq41}
\alpha_\Omega=\frac{4\pi\Gamma J}{U^3}\Big(U^2+\sqrt{U^4-J^2}\Big)\sqrt{5U^4+4U^2\sqrt{U^4-J^2}-J^2};
\end{equation}
it can be remarked that $\alpha_\Omega$ is well behaved and has no discontinuities, excluding $U=0$. This function is presented in Fig. \ref{fig08}. It can be noticed that $\alpha_\Omega$ is reduced when smaller values of $\Gamma$ are considered.

Noncommutative quantum corrected coefficient, $\alpha^\star_\Omega$ is given by,

\begin{displaymath}
\alpha^\star_\Omega=\frac{UJ\Big(U^2+\sqrt{U^4-J^2}\Big)\Big(4\pi\Gamma U^2+4\pi\Gamma\sqrt{U^4-J^2}-1\Big)^{3/2}}{-8\pi\Gamma U^8-U^6+4\pi\Gamma U^4J^2+2U^2J^2+\sqrt{U^4-J^2}(-8\pi\Gamma U^6-U^4+J^2)}
\end{displaymath}
\begin{equation}\label{eq42}
\times\Big[36\pi\Gamma U^6-5U^4-20\pi\Gamma U^2J^2+J^2-\sqrt{U^4-J^2}(-36\pi U^4+4\pi J^2+4U^2)\Big]^{1/2}.
\end{equation}
Analogously to other response functions, quantum corrected coefficient $\alpha^\star_\Omega$ have almost the same behavior than its Bekenstein-Hawking counterpart. If both curves are plotted together it is found that $\alpha_\Omega(U,J)>\alpha^\star_\Omega(U,J)$. 
As a summary, a comparison of thermodynamic properties between noncommutative Bekenstein--Hawking and quantum corrected entropies is presented in Table \ref{table01}.

\begin{table}[H]
	\caption{Comparison between thermodynamic properties of noncommutative Bekenstein--Hawking and noncommutative quantum corrected entropies.}
	\centering
	\label{table01}       
	\begin{tabular}{lll}
		\hline
		\textbf{Response functions} & \textbf{Equations of state} & \textbf{Fundamental relation}  \\
		\hline
		$C_J>C^\star_J$ & $T<T^\star$ & $S>S^\star$ \\
		$\chi_T<\chi^\star_T$ & $\Omega=\Omega^\star$  &\\
		$\chi_S=\chi^\star_S$ & &\\
		$C_\Omega<C^\star_\Omega$  & &\\ 
		$\alpha_\Omega>\alpha^\star_\Omega$  & &\\
		\hline
	\end{tabular}
\end{table}


In the following subsection, information provided by response functions  will be used to determine whether Kerr black holes are thermodynamically stable or not.  

\subsection{Thermodynamic stability and phase transition}\label{sec:3}

From the analysis performed on thermodynamic response functions for both, noncommutative $S$ and $S^\star$, an interesting result arises, specific heat capacity at constant angular momentum exhibits a singularity as showed in eqs. \eqref{eq23} and \eqref{eq28}. Discontinuity in $C_J$ for Bekenstein--Hawking entropy has been known for some time~\cite{Davies,Davies3}, and it is associated with a second--order (or continuous) phase transition. Expressions calculated from Bekenstein--Hawking entropy are more manageable and, for the sake of simplicity, the following analysis is performed considering only Bekenstein-Hawking thermodynamic properties. It is expected that results obtained with these considerations are prominently similar to the ones expected for quantum corrected properties.

Thermodynamic systems passing through a fist--order phase transition have physical states for which parts of the system are in different phases, or a phase coexistence, constituting a  series of not homogeneous states appearing below the critical point, where phase boundaries vanish. Often these states can be identified with the aid of thermodynamic diagrams, as $P$--$V$ diagrams for fluids. During phase transition equation of state remain constant, therefore a mechanical and thermal equilibrium exists~\cite{Huang,Atkins}.
Maxwell construction, is a correction to violation in van der Waals equation of the requirement to have a constant pressure with volume in isotherms of phase diagram during a first--order phase transition~\cite{Greiner}. It is helpful to find the critical point of a system in a first--order phase transition, if exists. 

For Kerr black holes isotherms in  $\Omega$-- $J$ plane must be analyzed. Criteria to find critical point is based on the pair of conjugate variables angular velocity and angular momentum, for which, the following requirements must be satisfied

\begin{equation}\label{eq43}
\Big(\frac{\partial\Omega}{\partial J}\Big)_{T_C}=0, \qquad \Big(\frac{\partial^2\Omega}{\partial J^2}\Big)_{T_C}=0;
\end{equation}
recalling eq. \eqref{eq29}, it implies that isothermic rotational susceptibility must be singular at this critical point $\chi_T\to\infty$.
As showed, $\chi_T$ do not have any divergence and is well behaved. Therefore, there is not critical point for Kerr Black holes. 
More evidence of this result can be found when constructing the isotherms in phase diagram in the plane $\Omega$--$J$. Changes in concavity of the curves are expected if the system pass through a first--order phase transition, this is the region of inhomogeneous states and it is commonly  named van der Waals loop, since were first observed for van der Waals equation.

\begin{figure}[t!]
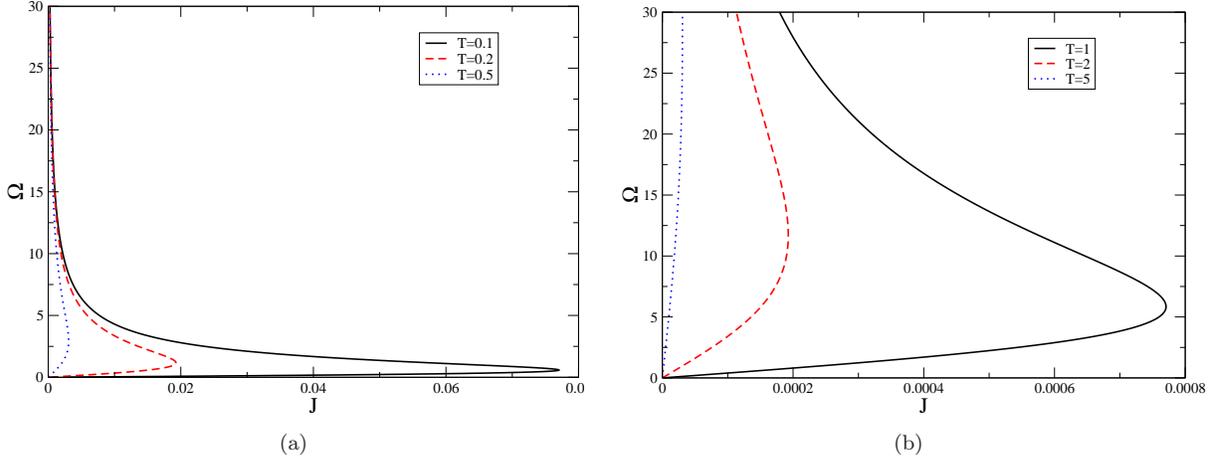

	\centering
	\subfigure[]{\includegraphics[width=0.45\textwidth]{Fig9a.eps}}
	\subfigure[]{\includegraphics[width=0.45\textwidth]{Fig9b.eps}}
	\caption{Isotherms in plane $\Omega$--$J$ for a Kerr black hole. Different temperatures were tested, exterior isotherm corresponds to the lower temperatures. Van der Waals loop do not appears in any of these isotherms. \label{fig09}}
\end{figure}

In order to construct the corresponding isotherm for noncommutative Bekenstein--Hawking Kerr black holes $\Omega=\Omega(J,T)$, it is easier to proceed from thermodynamic fundamental relation in energy representation $U(S,J)$~\cite{Escamilla1},

\begin{equation}\label{eq44}
U=\frac{1}{2}\sqrt{\frac{S}{\pi\Gamma}+\frac{4\pi\Gamma J^2}{S}}, \qquad \Omega=\frac{2\pi^{3/2}\Gamma J}{S\sqrt{\frac{S^2+4\pi^2\Gamma^2 J^2}{\Gamma S}}};
\end{equation}
using eq. (\ref{eq18}a) for temperature, it is straightforward to obtain:

\begin{equation}\label{eq45}
J=\Bigg(4\Omega^2\Bigg\{\Big[\Big(\frac{2\pi\Gamma T}{\Omega}\Big)^2+1\Big]^{3/4}+\Big(\frac{2\pi\Gamma T}{\Omega}\Big)\Big[\Big(\frac{2\pi\Gamma T}{\Omega}\Big)^2+1\Big]^{1/4}\Bigg\}\Bigg)^{-1}. 
\end{equation}
Inverse function $\Omega(T,J)$ can be estimated, and it is presented in Fig.~\ref{fig09} for different isotherms, showing commutative case. It was not possible to find an analytical expression for this relation. As noticed in this figure, for a given value of $J$ there are two corresponding values of $\Omega$, which can be interpreted as two possible cases for Kerr black holes, one of small mass and another with a larger one. When small temperatures are considered, it can be noticed that angular momentum have greater values available.
There is  no evidence of changes in concavity of the isotherms in plane $\Omega$--$J$, which implies that there is no van der Waals loop. The last piece of evidence is that if the analogous of the Maxwell construction for noncommutative Kerr black holes is tried to be performed, this procedure is not satisfied by any value in their domain, indicating once again, that there is not critical point.
Therefore, continuity of first derivatives along with the lack of critical point, indicates that Kerr black holes do not pass through a first--order phase transition.

Negative values exhibited by material properties are directly linked with thermodynamic stability of the system. Thermodynamic equilibrium states are characterized by an extremal principle, either maximal entropy or equivalently, a minimum in any other thermodynamic potential. In order to ensure that those potentials are stable, they must be concave functions of their natural variables. In particular, for Kerr black holes, Gibbs potential $G(T,\Omega)$ and Helmholtz free energy $F(T,J)$ must be concave functions of both temperature and angular velocity, temperature and angular momentum, respectively.
Using Legendre transformations for Kerr black holes~\cite{Escamilla1},

\begin{displaymath}
S=-\Big(\frac{\partial G}{\partial T}\Big)_\Omega=-\Big(\frac{\partial F}{\partial T}\Big)_J;
\end{displaymath}
concavity criteria requires that second derivatives satisfy the following relations~\cite{Stanley,Greiner}:

\begin{equation}\label{eq46}
\Big(\frac{\partial^2F}{\partial T^2}\Big)_J=-\Big(\frac{\partial S}{\partial T}\Big)_J=-\frac{1}{T}C_J\le0 \qquad (\text{For:} \ \Delta U\to0),
\end{equation}
\begin{equation}\label{eq47}
\Big(\frac{\partial^2G}{\partial \Omega^2}\Big)_T=-\Big(\frac{\partial J}{\partial \Omega}\Big)_T=-\chi_T\le0  \qquad (\text{For:} \ \Delta J\to0),
\end{equation}
\begin{equation}\label{eq48}
\Big(\frac{\partial^2G}{\partial T^2}\Big)_\Omega=-\Big(\frac{\partial S}{\partial T}\Big)_\Omega=-\frac{1}{T}C_\Omega\le0 \qquad (\text{For:} \ \Delta U\to0, \Delta J\to0 ).
\end{equation}
From the above relations and results found in section \ref{sec:2}, particularly eqs. \eqref{eq22}, \eqref{eq31} and \eqref{eq38}, it is evident that Kerr black holes have regions where thermodynamic states do not meet these requirements, since  $C_J$, $\chi_T$ and $C_\Omega$ are negative in those regions.

Geometric interpretation in the three dimensional thermodynamic space $S$--$U$--$J$ of eqs. (\ref{eq46})--(\ref{eq48}) can be found in the corresponding figures of each response function (see figures~\ref{fig04},~\ref{fig06} and~\ref{fig07}, respectively). For variations in internal energy, change of sign in $C_J$ implicates that noncommutative quantum corrected Kerr black holes are in weakly stable states (where some of the stability conditions are fulfilled, which are also known as metastable states), for low masses, becoming unstable at greater ones, as noticed in Fig. \ref{fig04}(a). For variations in angular momentum alone, as  showed in Fig. \ref{fig06}, isothermic rotational susceptibility becomes negative in the region above $J=0.68U^2$, namely, greater values of $J$, the system is also in weakly stable states for low values of $J$. When variations in both $U$ and $J$ are considered, noncommutative quantum corrected Kerr black holes are always unstable,  since $C_\Omega\le0$ in all its dominion, as presented in Fig. \ref{fig07}.

Although thermodynamic stability of extended Kerr black holes is not modified, varying the value of $\Gamma$ have a direct consequence on accessible weakly stable states, for example, increasing the region where $C_J$ is positive. As showed in Fig. \ref{fig05}, smaller values of noncommutativity parameter, force the system to exist in a larger set of metastable states.\\
Existence of thermodynamic instability and the divergence in $C_J$, reveals that the system goes through a series of metastable states, from a low mass black hole to a higher mass one, in analogy to other metastable phenomena as superheating or supercooling. Nevertheless, the lack of a microscopic description for black holes makes not possible to be sure that Kerr black holes pass through a continuous phase transition~\cite{Ruppeiner}. However, violation of stability criteria is a strong thermodynamic argument to support this hypothesis~\cite{Callen}.

\section{Conclusions}\label{sec:4}

An analysis on thermodynamic properties of noncommutative quantum corrected Kerr black holes using an approximate relation was presented. Although resulting expressions are mathematically more complicated, thermodynamic properties still retain the same functional behavior with respect to those calculated via Bekenstein--Hawking entropy.  
It was explicitly proven that Kerr black holes do not pass through a first--order phase transition, since the local criteria to find the critical point is not fulfilled for any value in the domain, corresponding isotherms do not exhibit van der Waals loops and the Maxwell construction cannot be obtained, all of which are characteristic of this kind of transition. Nonetheless, some second derivatives exhibit a change of sign which is an indication that those states are thermodynamically unstable. This instability and the nonexistence of a critical point suggest that the system goes through metastable states, from a low mass black hole to a high mass one, in a continuous phase transition.
Regarding the effective noncommutativity incorporated in the coordinates of minisuperspace, outside vicinity where $\Gamma\approx1$, changes introduced by this parameter over thermodynamic information of the system are relevant. In particular, it have a impact on stability of Kerr black holes, allowing the system to be thermodynamically metastable for a wider set of states.
Despite only an approximated expression was considered, it allowed us to study the effect of angular momentum for quantum noncommutative black holes. It would be interesting to have a complete description for noncommutative rotating black holes, in order to compared with the results presented in ths work, in particular, those related with thermodynamic stability.

\vspace{6pt} 


\acknowledgments{Lenin F. Escamilla-Herrera wants to thank Programa de Mejoramiento del Profesorado (PROMEP, SEP), grant DSA/103.5/14/8614, for the support in the realization of this work.\\ 
J. Torres-Arenas thanks the financial support from the Consejo Nacional de Ciencia y Tecnolog\'ia (CONACyT, M\'exico), project number 152684; and Universidad de Guanajuato, project  740/2016 Convocatoria Institucional de Investigaci\'on Cient\'ifica.\\
	}
    
	\renewcommand\bibname{References}


\begin{thebibliography}{999}
		\bibitem{Hawking} Hawking, S.W. Gravitational Radiation from Colliding Black Holes. {\em Phys. Rev. Lett.} {\bf 1971}, {\em 26},  1344.
		\bibitem{Hawking2} Hawking, S.W. Black holes and thermodynamics. {\em Phys. Rev. D} {\bf 1976}, {\em 13}, 191.
		\bibitem{Bekenstein} Bekenstein,  J.D. Black Holes and Entropy. {\em Phys. Rev. D} {\bf 1973}, {\em 7}, 2333.
		\bibitem{Bekenstein1} Bekenstein, J.D. Generalized second law of thermodynamics in black-hole physics. {\em Phys. Rev. D} {\bf 1974}, {\em 9}, 3292.
		\bibitem{Chamseddine} Chamseddine, A.H. Deforming Einstein's gravity {\em Phys. Lett. B} {\bf 2001} {\em 504}, 33.
		\bibitem{Garcia-Compean} Garc\'ia-Compe\'an, H.; Obreg\'on, O.; Ram\'irez, C. Noncommutative Quantum Cosmology. {\em Phys. Rev. Lett.} {\bf 2002}, {\em 88}, 161301.
		\bibitem{Gibbons} Gibbons, G.W., Hawking, S.W. Cosmological event horizons, thermodynamics, and particle creation. {\em Phys. Rev. D} {\bf 1977}, {\em 15}, 2738-275.
		\bibitem{Davies3} Davies P.C.W., Thermodynamic phase transitions of Kerr-Newman black holes in de Sitter space. {\em Class. Quantum Grav.} {\bf 1989}, {\em 6}, 1909.
		\bibitem{Chamblin} Chamblin, A.; Emparan, R.; Johnson, C.V.; Myers, R.C. Charged AdS black holes and catastrophic holography. {\em Phys. Rev. D} {\bf 1999}, {\em 60}, 064018.
		\bibitem{Hawking-Page} Hawking, S.W.; Page, D.N. Thermodynamics of black holes in anti-de Sitter space. {\em Comm. Math. Phys.} {\bf 1983}, {\em 87}, 577-588.
		\bibitem{Price} Price; R.H. Nonspherical perturbations of relativistic gravitational collapse. 1. scalar and gravitational perturbations. {\em Phys. Rev. D} {\bf 1972}, {\em 5}, 2419.
		\bibitem{Gubser} Gubser, S.S.; Mitra I. The evolution of unstable black holes in anti-de Sitter space. {\em JHEP} {\bf 2001}, {\em 0108}, 018. 
		\bibitem{Wald} Hollands, S.; Wald, R.M. Stability of Black Holes and Black Branes. {\em Comm. Math. Phys.} {\bf 2013}, {\em 321}, 629-680.
		\bibitem{Davies} Davies, P.C.W. The thermodynamics of black holes. {\em Proc. R. Soc. Lond. A} {\bf 1977}, {\em 353}, 499.
		\bibitem{Martinez} Martinez, E.A. The postulates of gravitational thermodynamics. {\em Phys. Rev. D} {\bf 1996}, {\em 54}, 6302.
		\bibitem{Escamilla1} Escamilla, L.; Torres-Arenas, J. Thermodynamic response functions and Maxwell relations for a Kerr black hole. {\em Rev. Mex. Fis.} {\bf 2014}, {\em 60},  59-68.
		\bibitem{Hawking4} Hawking; S.W., Gravitational Radiation from Colliding Black Holes, {\em Phys. Rev. Lett.} {\bf 1971}, {\em 26} 1344.
		\bibitem{Keeler} Keeler, C.; Larsen, F.; Lisb\~ao, P. Logarithmic corrections to $\mathcal{N}\ge 2$ black hole entropy. {\em Phys. Rev. D} {\bf 2014}, {\em 90}, 043011.
		\bibitem{Sen} Sen, A. Logarithmic corrections to Schwarzschild and other non-extremal black hole entropy in different dimensions. \textit{\em J. High Energy Phys.} {\bf 2013}, {\em 04, 156}.
		\bibitem{Obregon} L\'opez-Dom\'inguez, J.C.; Obreg\'on, O.; Sabido, M.; Ram\'irez, C. Towards Noncommutative Quantum Black Holes. {\em Phys. Rev. D} {\bf 2006}, {\em 74}, 084024.
		\bibitem{Sen2} Sen, A. Logarithmic corrections to $\mathcal{N}\ge 2$ black hole entropy: an infrared window into the microstates. {\em Gen. Relativ. Gravit.} {\bf 2012}, {\em 44}, 1207-1266.
		\bibitem{Banerjee} Banerjee, S.; Gupta, R.K.; Mandal I.; Sen, A. Logarithmic corrections to and black hole entropy: a one loop test of quantum gravity. {\em J. High Energy Phys.} {\bf 2011}, {\em 03, 143}.
		\bibitem{Carlip} Carlip, S. Logarithmic corrections to black hole entropy, from the Cardy formula. {\em Class. Quantum Grav.} {\bf 2000}, {\em 17}, 4175.
		\bibitem{Kantowski} Kantowski, R.; Sachs, R.K. Some Spatially Homogeneous Anisotropic Relativistic Cosmological Models. {\em J. Math. Phys.} {\bf 1966}, {\em 7}, 443.
		\bibitem{Gamboa} Gamboa, J.; Loewe, M.; Rojas, J.C. Noncommutative quantum mechanics. {\em Phys. Rev. D} {\bf 2011}, {\em 64}, 067901.
		\bibitem{Feynman} Feynman, R.P.; Hibbs, A.R. In {\em Quantum Mechanics and Path Integrals}; 1st edn.; McGraw-Hill: New York, US, 1965.
		\bibitem{Aoki} Aoki, H.; Ishibashi, N.; Iso, S.; Kawai, H.; Kitazawa Y.; Tada, T. Noncommutative Yang-Mills in IIB Matrix Model. {\em Nucl. Phys. B} {\bf 2000}, {\em 565}, 176.
		\bibitem{Joby} P.K., Joby; Chingangbam P.; Das, S. Constraint on noncommutative spacetime from PLANCK data. {\em Phys. Rev. D} {\bf 2015}, {\em 91},  083503.
		\bibitem{Sabido} Guzm\'an, W.; Sabido, M.; Socorro, J. On noncommutative minisuperspace and the Friedmann equations. {\em Phys. Lett. B} {\bf 2011}, {\em 697}, 271.
		\bibitem{Callen} Callen, H.B. In {\em Thermodynamics and an Introduction to Thermo-statistics}; 2nd edn.; John Willey \& Sons: Singapore, 1985; pp. 35, 183, 215.
		\bibitem{Stanley} Stanley, H.E. In {\em Introduction to Phase Transitions and Critical Phenomena}; 1st edn.; Oxford University Press: London, UK, 1971; pp. 34, 29.
		\bibitem{Bagher} Poshteh, M.B.J.; Mirza,  B.; Z. Sherkatghanad, Z. Phase transition, critical behavior, and critical exponents of Myers-Perry black hole. {\em Phys. Rev. D} {\bf 2013}, {\em 88}, 024005.
		\bibitem{Banerjee2} Banerjee, R.; Roychowdhury, D. Thermodynamics of phase transition in higher dimensional AdS black holes. {\em J. High Energy Phys.} {\bf 2011}, {\em 1111}, 004.
		\bibitem{Mansoori} Mansoori S.A.H.; Mirza, B. Correspondence of phase transition points and singularities of thermodynamic geometry of black holes. {\em Eur. Phys. J. C} {\bf 2014}, {\em 74}, 2681.
		\bibitem{Lala} Lala, A.; Roychowdhury, D. Ehrenfest’s scheme and thermodynamic geometry in Born-Infeld AdS black holes. {\em Phys. Rev. D} {\bf 2012}, {\em 86}, 084027.
		\bibitem{Huang} Huang, K. In {\em Statistical Mechanics}; 2nd edn.; John Willey and Sons: New York, US, 1987; pp 32, 40.
		\bibitem{Atkins} Atkins, P.; de Paula, J. In {\em Physical Chemistry}; 8th edn.; W.H. Freeman and Company: New York, US, 2006; pp 21. 
		\bibitem{Greiner} Greiner, W.; L. Neise L.; St\"ocker, H. In {\em Thermodynamics and Statistical Mechanics}, 1st edn.; Springer-Verlag: New York, US, 1995; pp 68, 119. 
		\bibitem{Ruppeiner} Ruppeiner, G. Stability and fluctuations in black hole thermodynamics. \textit{Phys. Rev. D} \textbf{75}, 024037 (2007).
\end{thebibliography}
\end{document}